\documentclass[aps,twocolumn,nofootinbib,preprintnumbers,prl,10pt]{revtex4-2}
\usepackage[utf8]{inputenc}

\usepackage[svgnames]{xcolor}
\usepackage[most]{tcolorbox}
\usepackage{caption}
\tcbset{colback=PaleGreen!0!white, colframe=NavyBlue, 
	highlight math style= {enhanced, 
		colframe=red,colback=red!10!white,boxsep=0pt}
}
\usepackage{relsize}

\usepackage{amsmath}
\usepackage[bottom]{footmisc}
\usepackage{braket}
\usepackage{graphicx}
\usepackage{mwe}
\usepackage[section]{placeins}
\usepackage{framed}
\usepackage{csquotes}
\usepackage{tikz}
\usetikzlibrary{decorations.markings}
\usetikzlibrary{decorations.pathmorphing}
\definecolor{shadecolor}{rgb}{0.90,0.90,0.90}
\usepackage{hyperref}
\usepackage{bbold}
\usepackage{subcaption}
\usepackage{pifont}
\usepackage{setspace}
\usepackage{amsmath, amssymb, amsthm, float, graphicx,amsfonts}
\usepackage{amsthm}
\usepackage{array, boldline, makecell, booktabs}

\theoremstyle{definition}

\hypersetup{colorlinks=true, linkcolor=DarkRed, citecolor=DarkRed,urlcolor=Indigo,linktocpage}
\def\beq{\begin{eqnarray}}\def\eeq{\end{eqnarray}}
\def\be{\begin{equation}}\def\ee{\end{equation}}
\def\bs{\begin{split}}\def\es{\end{split}}

\usepackage{bm}

\usepackage{mathrsfs}
\usepackage{slashed}

\begin{document}

\title{\bf Fuzzballs and Random Matrices}

\author{\!\!\!\! Suman DAS$^{a}$, Sumit K. GARG$^{b}$, Chethan KRISHNAN$^{c}$, Arnab KUNDU$^{a}$\\ ~~~~\\
\it ${^a}$Theory Division, Saha Institute of Nuclear Physics,\\
A CI of Homi Bhabha National Institute,\\
1/AF, Bidhannagar, Kolkata 700064, India \\
{\rm Email: \textsf{suman.das@saha.ac.in, arnab.kundu@saha.ac.in}}\\
\it ${^b}$Manipal Centre for Natural Sciences, \\
Manipal Academy of Higher Education, \\
Dr. T.M.A. Pai Planetarium Building, \\ 
Manipal-576104, Karnataka, India\\
{\rm Email: \textsf{sumit.kumar@manipal.edu}}\\
\it ${^c}$Center for High Energy Physics,
\it Indian Institute of Science,\\ \it C.V. Raman Road, Bangalore 560012, India. \\
{\rm Email: \textsf{chethan.krishnan@gmail.com}}}

\begin{abstract}{Black holes are believed to have 
the fast scrambling properties of random matrices. If the fuzzball proposal is to be a viable model for quantum black holes, it should reproduce this expectation. This is considered challenging, because it is natural for the modes on a fuzzball microstate to follow Poisson statistics. In a previous paper, we noted a potential loophole here, thanks to  the modes depending not just on the $n$-quantum number, but also on the $J$-quantum numbers of the compact dimensions. For a free scalar field $\phi$, by imposing a Dirichlet boundary condition $\phi=0$  at the stretched horizon, we showed that this $J$-dependence leads to a linear ramp in the Spectral Form Factor (SFF). Despite this, the status of level repulsion remained mysterious.  In this letter, motivated by the profile functions of BPS fuzzballs, we consider a generic profile $\phi = \phi_0(\theta)$ instead of $\phi=0$ at the stretched horizon. For various notions of genericity (eg. when the Fourier coefficients of $\phi_0(\theta)$ are suitably Gaussian distributed), we find that the $J$-dependence of the spectrum exhibits striking evidence of level repulsion, along with the linear ramp. We also find that varying the profile leads to natural interpolations between Poisson and Wigner-Dyson(WD)-like spectra. The linear ramp in our previous work can be understood as arising via an extreme version of level repulsion in such a limiting spectrum. We also explain how the stretched horizon/fuzzball is different in these aspects from simply putting a cut-off in flat space or AdS (ie., without a horizon). 









}
\end{abstract}
\maketitle

{\noindent \bf Introduction}: The quest for an understanding of quantum black holes has been one of the engines driving research in quantum gravity in the last half century. In particular, the recent revival of the black hole information paradox \cite{Hawking, Page} due to the works of Mathur \cite{Mathur} and AMPS \cite{AMPS} has raised questions about the smoothness of the horizon which are still not fully settled. 

In the context of holography/string theory, there are two broad lines along which work on quantum black holes has progressed. The first approach, which we will call the {\em semi-classical approach} following \cite{synthfuzz}, is built on insights from bulk (often Euclidean) effective field theory, toy models of 2D gravity, and holographic entanglement entropy. Considerable intuition has been gleaned about the quantum nature of black holes from this approach (eg. \cite{Sekino, Butterfly, MSS, RT, Engelhardt-Wall}) with the crowning achievement being a semi-classical reproduction of the Page curve \cite{Penington, Almheiri}. Despite this, the precise status of detailed unitarity and smoothness are still unclear from this perspective, because the calculation is fundamentally Euclidean. The second line of approach is the {\em fuzzball program} of Mathur and others which argues that black hole microstates cap off smoothly before the horizon. In our opinion, the operational meaning of this bulk statement in the full quantum setting is not yet completely clear. But the mere existence of large classes of such solutions \cite{Lunin,  Rychkov, KST, Avinash, superstrata, MathurReview, BenaReview, Masaki} in the supergravity limit of stringy BPS black holes is surprising. In conventional general relativity, they would not exist thanks to the no hair theorems. See \cite{synthfuzz} for a more detailed discussion of the pros and cons of the two approaches. 

It was suggested in \cite{synthfuzz} that one way to make progress may be to try and reproduce {\em general} lessons of the semi-classical approach, from fuzzball-motivated considerations. The hope is that since many of these expectations are generic, this may teach us something about how to think about quantum fuzzballs at finite temperature even though constructing explicit solutions is possible only in the supergravity BPS limit. Conversely, if realizing these lesson from fuzzball-motivated ideas is impossible or highly contrived, that could be viewed as an argument against the fuzzball program. 

A particularly sharp setting in which one could explore this tension is in the expectation that black holes are fast scramblers \cite{Sekino}, and that they exhibit dynamical features of random matrices \cite{Cotler}. A linear ramp \cite{linear} in the spectral form factor (SFF) and repulsion in the level spacing distribution (LSD), are viewed as indicators of chaos in random matrix theory (RMT) \cite{SFF}. However, these RMT signatures are generally thought to be challenging to realize from the fuzzball paradigm, see eg. \cite{Martinec} --  we expect capped geometries to have roughly evenly spaced levels, in loose analogy with the standing waves of a cylindrical column. This makes conventional level repulsion and the linear ramp, difficult to understand from the fuzzball perspective. Note also that simply declaring that the black hole is an {\em ensemble} of such spectra, does not solve the problem \cite{ensemble} -- While this will certainly allow a richer set of level spacings in the collective spectrum, there is still no mechanism to ensure level repulsion \cite{SFF-LSD}. Instead, an ensemble of fuzzballs will give rise to Poisson statistics, just as an ensemble of simple harmonic oscillators (SHO) would \cite{Appendix}.

These expectations are reasonable, but they are also difficult to test. This is because solving wave equations in generic fuzzball microstate geometries is both difficult (because the metric is complicated) and not immediately useful (because explicit metrics in BPS cases are at zero temperature). Exploiting the fact that the questions we wish to tackle are generic, in \cite{synthfuzz} it was suggested that one may be able to make progress by studying a black hole at finite temperature with a stretched horizon. In particular, the normal modes of a scalar field were studied in \cite{synthfuzz}, by computing the spectrum of modes that result from a $\phi=0$ boundary condition at the stretched horizon. The results of \cite{synthfuzz} showed that the expectations listed in the previous paragraph have a major caveat, they are true {\em only if one ignored the dependence of the spectrum on the angular quantum numbers of the compact dimensions}. Unlike the dependence on the principal $n$-quantum number, the dependence on the $J$-quantum numbers was found {\em not} to be (approximately) linear. Instead there was a quasi-degeneracy of levels as a function of $J$ for moderately large $J$. Most strikingly, it was found that the SFF computed from the spectrum showed very clear evidence of a linear ramp, even though conventional level repulsion was not present in the $J$-direction \cite{spectrum}. It should be emphasized here that this is the only case in the literature that we are aware of, where a linear ramp in the SFF exists without an underlying RMT spectrum with Wigner-Dyson (WD) level spacing \cite{rampy-spectra}. 

While the results of \cite{synthfuzz} were a tantalizing hint of RMT behavior in fuzzballs, a coherent understanding of them could not be found. In particular, the {\em presence} of a linear ramp together with the {\em absence} of conventional level repulsion, made a compelling interpretation impossible. The purpose of this letter, is to shed some light on this mysterious state of affairs. We will place the results of \cite{synthfuzz} in context by finding a more general calculation that can interpolate between Poisson and RMT-like spectra. The idea (at least in hindsight) is extremely simple, and motivated by the fact that the known BPS fuzzball solutions \cite{Lunin, KST, superstrata} are described by profile functions that are supposed to capture the fluctuations of the cap. This suggests that a natural generalization of our simple $\phi=0$ boundary conditions of \cite{synthfuzz} is to consider a generic profile $\phi=\phi_0(\theta)$ at the stretched horizon, where $\theta$ is a mnemonic for the angular directions of the metric. In this paper, we will consider profiles of this type, where ``genericity'' will be implemented via choosing Fourier coefficients of $\phi_0(\theta)$ from suitable random distributions. This is a natural implementation of the intuitive notion of ``fluctuation at the horizon''. Remarkably, in this very natural set up, we see both level repulsion as well as the linear ramp. By tuning the variance of the distribution from which $\phi_0(\theta)$ is chosen, we show that the LSD can interpolate from Poisson to WD-like spectra. In particular, as the variance collapses to zero and the boundary condition reduces to $\phi=0$, we find that the LSD collapses to a very sharp (almost delta-function-like) peak, as found in  \cite{synthfuzz}. It was speculated in \cite{synthfuzz} that this should be viewed as an ``extreme'' version of level-repulsion, and our present paper clarifies the precise sense in which this is true. Conversely, as the variance is steadily increased, the LSD transitions from ``extreme" to conventional Wigner-Dyson spectra and eventually to Poisson \cite{LSR0}. 

Our results demonstrate that fuzzball/stretched horizon modes can exhibit the spectral features of RMT and late time chaos. We emphasize that this is a {\em bulk} calculation of RMT behavior. The expectation of RMT behavior and eigenstate thermalization in black hole microstates is natural in the {\em dual} holographic theory, because it is strongly coupled. This has been explicitly demonstrated in the setting of toy dual theories like SYK and tensor models \cite{SYK}. From the bulk however, while early time chaos is captured by out-of-time-ordered correlators \cite{Butterfly, MSS}, late-time chaos as captured by level repulsion and discreteness of the spectrum are very difficult to understand. Fuzzballs can exhibit discreteness in the spectrum trivially, by virtue of the fact that they do not have a horizon. On the other hand as we noted earlier, the origin of RMT behavior from fuzzballs is supposedly non-trivial to arrange. Our results show on the contrary, that there are generic bulk mechanisms that can enable fuzzballs to capture RMT features.

In the following section, we will present our main results while relegating the technical details to various Supplementary Material. To give further confidence that these results really do have to do with the magic of black holes and horizons, we will also discuss some examples where there are no horizons. Putting a cut-off in such geometries leads to major qualitative differences from stretched horizons, which we elaborate. In the Conclusions section we review and emphasize the salient points of our results and extract some lessons. Some related further observations and comments about  future directions \cite{GRMT}, as well as various technical details, are presented in various Supplementary Material.

{\noindent \bf{Main Results}}: We will solve the massless scalar field equation in a black hole geometry with a stretched horizon, while demanding the boundary condition $\phi = \phi_0(\theta)$ at the stretched horizon. We will do this for the BTZ black hole as well as for the Rindler wedge  (times a compact space); these were the two cases studied in detail in \cite{synthfuzz}. The primary virtue of these choices is that the wave equation is solvable in terms of well-known special functions. We will see that the resulting physics is identical in both cases, and we do not expect qualitative changes in our conclusions for other black holes, in 2+1 dimensions and higher. 

The details of the wave equations and how we obtain the normal modes for a general stretched horizon profile  are presented in the Supplementary Material. The scalar field boundary condition profile can be described in terms of its Fourier coefficients. We will choose each of these Fourier coefficients randomly from a suitable Gaussian distribution (see the discussion in the Supplementary Material, for details on how this is done). This means that there are two choices we need to make in order to fully define the problem -- the mean and the variance of this Gaussian distribution \cite{uniform}. To make sure that the Fourier series sum converges and leads to a well-defined profile, we will also cut-off the sum at some $J$. This should be compared to the cut-off in $J$ that is required to define the SFF \cite{synthfuzz}. It turns out that the mean and the variance have a heuristic (but suggestive) interpretation in terms of the location and the fluctuations of the stretched horizon, see again the Supplementary Material. To have a natural interpretation as the stretched horizon at a Planck length, we will take the mean to be very large in tortoise coordinates (and therefore close to the horizon). Note that since we are working with a fixed background geometry, the Planck length is an arbitrary choice. 

Our conclusions are entirely analogous for both BTZ and Rindler, so we will discuss BTZ here for concreteness; see Figures. 
\begin{figure}[h]
    \centering
    \includegraphics[width=.47\textwidth]{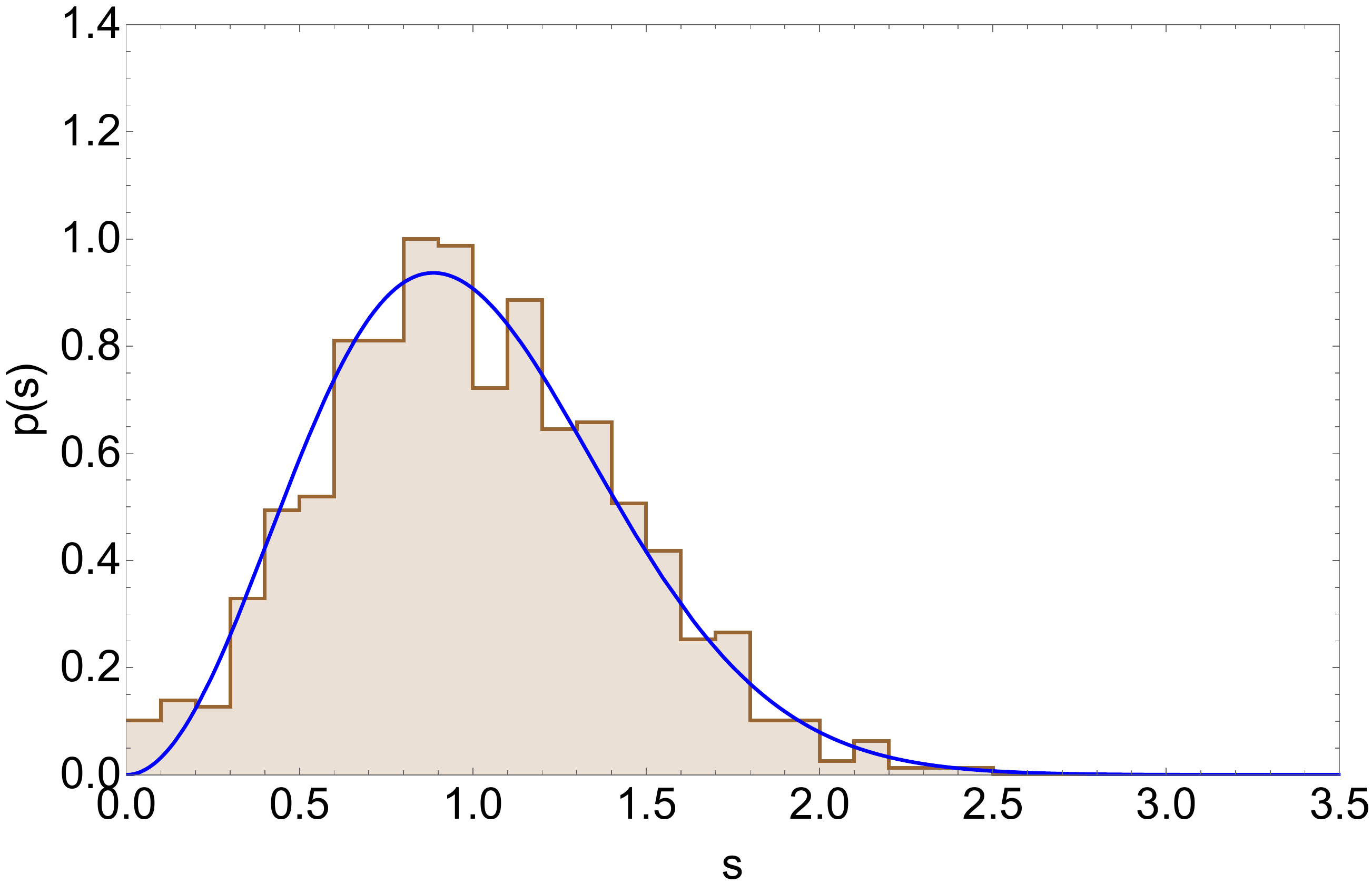}
   \caption{LSD for BTZ black hole normal modes  $\omega(n=1,J)$, with $\langle \lambda \rangle=-10^3$, $J_{max}=800$ and $\sigma_{\lambda_J}=\sigma_0/J$ with $\sigma_0=0.3$ . Supplementary Material contains definitions and explanations of the notation. The blue curve is the GUE level spacing curve.}
    \label{btz_main_lsd}
\end{figure}
\begin{figure}[h]
    \centering
    \includegraphics[width=.47\textwidth]{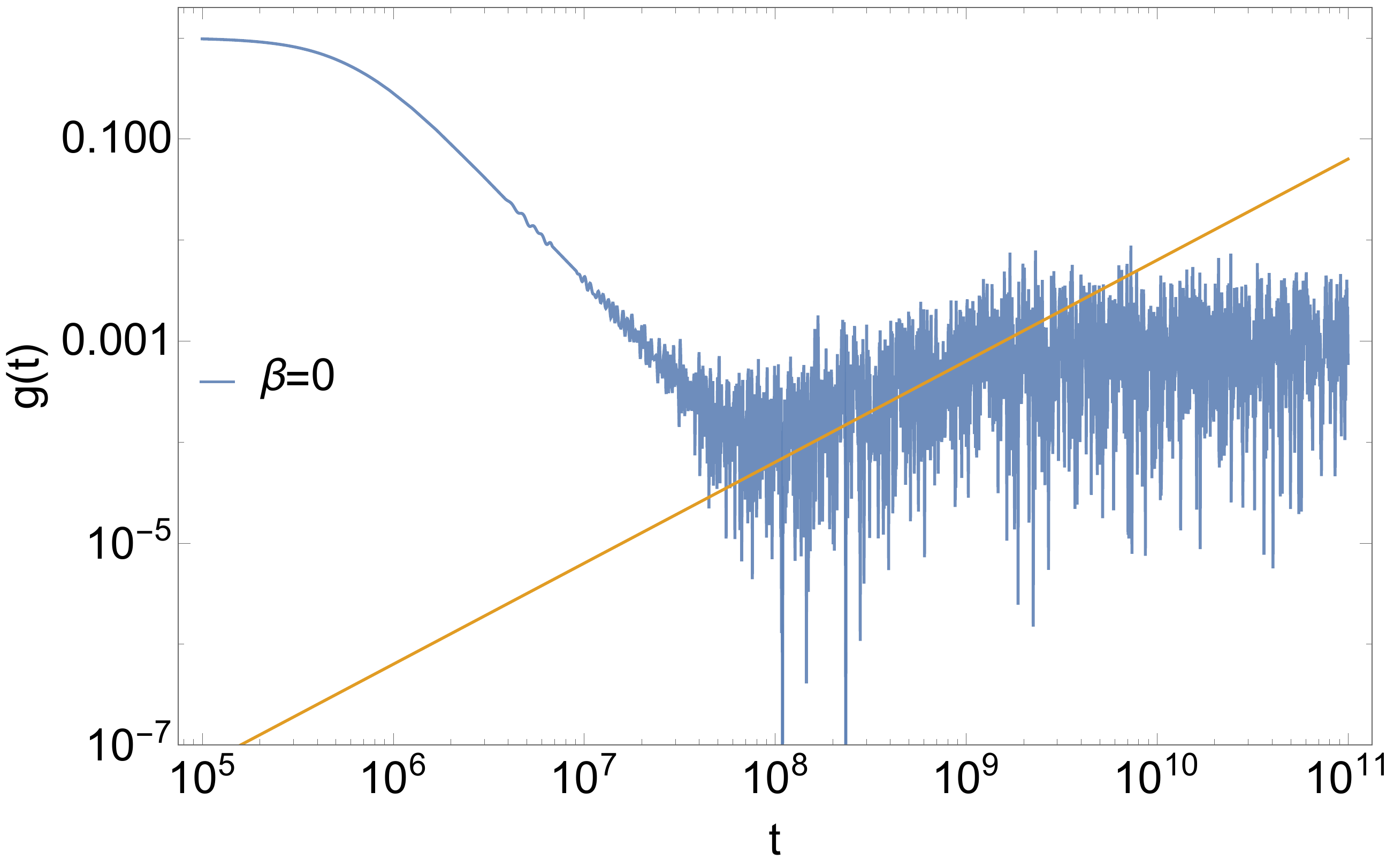}
   \caption{SFF for BTZ black hole normal modes; same parameters as above. The slope of the line is unity. Together these two figures (and the many others in the Supplementary Material) show that we can get {\em both} the linear ramp as well as level repulsion from ``synthetic'' fuzzball normal modes.}
    \label{btz_main_sff}
\end{figure}
More plots and discussions are provided in  the Supplementary Material.  To summarize -- Our results for the SFF and the LSD reduce to those of \cite{synthfuzz} when the variance is zero; the SFF has a linear ramp, but the LSD is of the ``extreme" delta function-like  form. But remarkably, for small but non-zero choices of the variance, one finds LSDs that fit Wigner-Dyson \cite{WhichWD}, while the linear ramp remains intact. Finally, as the variance becomes large, the LSD reduces to the Poisson form and the ramp goes away. 

These results are qualitatively different from corresponding results in a geometry where a cut-off is introduced {\em without} a horizon. To demonstrate this, we also study flat space and AdS with a cut-off. Once again the details of the computation and plots are presented in the Supplementary Material. In flat space, we find that there is never a ramp of slope $\sim 1$, but for moderate variances, there is a clear {\em non}-linear ramp of slope $\sim 1.9$. 
The level-spacing distribution when there is no variance is again a strongly peaked delta-function-like form. But the origin of this fact has a simple (and less interesting) understanding, as opposed to when there was a horizon. In flat space the levels are roughly evenly spaced and therefore the spectrum is analogous to that of an SHO (which also has a delta function LSD, even though it is the farthest thing from RMT). 
Indeed, we have directly checked that the SFF of an SHO with a small amount of noise added to its energy levels, reproduces precisely the non-linear ramp of slope $\sim 1.9$ we noted above. This, and some interesting related results are presented in some of the Supplementary Material and a follow-up paper \cite{GRMT}. The bottom line is that the linearity of the ramp is lost when we simply put a cut-off in flat space as opposed to at a stretched horizon. Loosely similar statements hold in AdS as well. We will suppress the details, except to mention that one has to take care of two separate cases. One where the cut-off size is much larger than the AdS scale, and another where it is much smaller. The latter turns out to yield a discussion identical to the flat space case above (as expected). In the former case, there is no well-defined constant slope ramp at all in the log-log plot, so it will not be of interest to us here.

A second distinction between the modes of a horizon-less cut-off and a stretched horizon is that the variance one introduces in the former case can  heuristically be interpreted as due to {\em macroscopic} fluctuations at the cut-off. In the stretched horizon case, the fluctuations are in the tortoise coordinate, and therefore have a natural interpretation as Planckian suppressed. This is again very natural from the membrane paradigm/fuzzball perspective. These matters are discussed in detail in the Supplementary Material.

{\noindent \bf{Conclusions}}: Our goal in  \cite{synthfuzz} and this paper has been to see whether the fuzzball/stretched horizon paradigm can be useful for reproducing some of the successes of the semi-classical approach to quantum black holes. As pointed out in \cite{synthfuzz}, both approaches have produced interesting results, yet major open problems remain. While the stretched horizon/fuzzball will trivially get rid of some aspects of the information paradox, finding hints of RMT behavior is considered challenging. 

We demonstrated that we can find both the linear ramp and conventional level repulsion (as well as RMT level spacing ratios) from a stretched horizon. The linear ramp is a direct consequence of a cut-off near the horizon. In a cut-off geometry without a horizon, the linear ramp never exists, and a non-linear ramp when it exists, can be understood as related to an SHO spectrum with noise. We also found that conventional level repulsion is easy to realize, 
by simply incorporating angular dependence in the boundary condition. This is interesting, because such angle-dependence is generic in BPS fuzzball microstates. 

The existence of the linear ramp is usually taken as a signature of strong chaos. Finding the linear ramp in our previous paper \cite{synthfuzz} was encouraging, but the absence of conventional level repulsion made the result puzzling. 
But given the ramp, it is natural to suspect that some small perturbation may be able to produce the nearest-neighbor correlations \cite{Hanada-comment} as well. The challenge was to identify the right kind of perturbation. The fluctuations at the stretched horizon that we have included in this paper can be viewed as a natural candidate for such a small perturbation. The variance in the Fourier modes of the fluctuation profile leads to a small noise in the spectrum, which leads to the requisite spread in the LSD.

Our results also strengthen the case that level repulsion is a weaker hint of chaos than the linear ramp. This is because it is only sensitive to nearest neighbor physics. We explicitly demonstrate this using the example of the SHO in the Supplementary Material, where it is shown that adding a small amount of noise to the SHO energy levels is sufficient to produce conventional WD-like LSD plots. But this is {\em not} sufficient to produce the linear ramp, which is sensitive to long range correlations within the spectrum. This again ties nicely with the observation that the linear ramp is present only when the cut-off is near the black hole horizon, while level repulsion can be realized in a cut-off geometry with or without a horizon as long as we are working with a fluctuating profile \cite{fluc}. The SFFs of horizonless cases with moderate variance have a power law ramp of slope $\sim 1.9$ -- This is the same as that of a moderately noisy SHO.

A natural proposal that ties together our observations then, is as follows -- Signatures of robust chaos (as captured by the linear ramp) emerge when we consider a stretched horizon close to the black hole.  Such signatures are not present when the cut off is in empty space or far from the horizon. These statements are independent of the profile choices at the cut-off. But the profiles do play a role, when we are discussing nearest neighbor physics and level repulsion in the system. A profile with non-vanishing variance can lead to nearest-neighbor level repulsion both with or without a horizon, but the natural length scale associated to the variance has to be macroscopic for this to happen in a horizonless geometry. In other words, even if we allow macroscopic fluctuations, we can at best see nearest neighbor effects in a horizonless geometry with a cut-off. On the contrary, stretched horizon/fuzzball modes automatically carry signatures of robust chaos and a linear ramp, with or without a non-trivial profile. If the profile is generic in the sense of having a small non-zero variance, they reproduce the correct nearest neighbor effects as well.

Semi-classical bulk calculations involving replica wormholes (and implicitly, ensemble averages) are known to produce a smooth linear ramp without fluctuations. The challenge for quantum gravity is to reproduce a linear ramp without any ensemble average from a single microstate, and {\em with} fluctuations. Our calculation, despite its simplicity has reproduced both these features. This may seem surprising because our set up is superficially (semi-)classical. But this is misleading -- The boundary conditions we are imposing at the stretched horizon, while technically simple, are conceptually highly non-trivial from the dual CFT. It is clearly of interest to understand this boundary condition better from an intrinsically CFT perspective.

It may be useful to re-visit the many questions about (quantum) black holes at finite temperature, armed with the perspectives we have added in this paper. In this section, we have only emphasized black hole physics. A more detailed discussion of open questions and questions more intrinsic to RMT physics are presented in the Supplementary Material. 

\section*{Acknowledgments}
 
We thank A. Preetham Kumar for crucial contributions in our previous collaboration \cite{synthfuzz}, and Masanori Hanada, Romesh Kaul, Alok Laddha, R. Loganayagam,  Ayan Mukhopadhyay, Onkar Parrikar, Ashoke Sen, Kostas Skenderis and Amitabh Virmani for discussions and/or correspondence.

\onecolumngrid

\newpage
{\begin{center}\bf \Large{Supplementary material}\end{center} }

\section{Case Study: BTZ}

As in \cite{synthfuzz}, we will start by considering a scalar field $\Phi$ of mass $m$ in the BTZ background,
\begin{equation}\label{BTZ_metric}
    ds^2=-\frac{(r^2-r_h^2)}{L^2}dt^2+\frac{L^2}{(r^2-r_h^2)}dr^2+r^2 d\psi^2
\end{equation}
with $-\infty<t<\infty$, $0<r<\infty$ and $0\le \psi< 2\pi$. In \cite{synthfuzz} we fixed units by setting $L=1$ and worked with the numerical choice $r_h=1$ from the outset. Here, we will present the more general expressions because it is useful in comparisons with cut-off empty space. The new boundary conditions and the corresponding results/plots start only after \eqref{PQ2}. So a reader who is familiar with the results of \cite{synthfuzz} and is willing to believe the slightly more general expressions we present here, can skip directly to the discussion after \eqref{PQ2}. 

The wave equation
\begin{equation}\label{eom1}
    \Box \Phi\equiv \frac{1}{\sqrt{|g|}}\partial_{\mu}\left(\sqrt{|g|}\partial^{\mu}\Phi\right)=m^2 \Phi
\end{equation}
can be solved by writing 
\begin{equation}
   \Phi = \frac{1}{\sqrt{r}} \sum_{\omega, J} e^{-i\omega t}  e^{i J\psi} \phi_{\omega, J}(r) \label{scalarmodes}
\end{equation}
with integer $J$. The radial part of \eqref{eom1} satisfies,

\begin{equation}
    (r^2-r_h^2)^2 \phi^{''}_{\omega, J}(r)+2 r(r^2-r_h^2) \phi^{'}_{\omega, J}(r)+   \omega^2 L^4 \phi_{\omega, J} (r)
 -V_J(r)\phi_{\omega, J}(r) =0 \label{radial_eom}
\end{equation}
where \begin{equation}
    V(r)=(r^2-r_h^2)\left[\frac{1}{r^2}\left(J^2 L^2+\frac{r_h^2}{4}\right)+\nu^2-\frac{1}{4} \right], \hspace{2cm} \nu^2=1+m^2 L^2.
\end{equation}
We will generally work with the massless case, $\nu=1$. The solution\footnote{We will work with the massless scalar and the $J=0$ mode needs special treatment. See footnote 13 of \cite{synthfuzz}.}
of this is given in terms of hypergeometric functions as
\begin{equation}\label{sol1}
    \phi(r)=\left(r\right)^{\frac{1}{2}-\frac{i J L}{r_h}}\left(r^2-r_h^2\right)^{-\frac{i \omega L^2}{2 r_h}}\left(e^{-\frac{\pi J L}{r_h}}\left(\frac{r}{r_h}\right)^{\frac{2i J L}{r_h}}C_2  H\left(r\right)+C_1  G\left(r\right) \right),
\end{equation}
where we are suppressing the subscripts $\omega, J$ on the LHS as well as on $C_1$ and $C_2$. Here,  
\begin{align}
    G\left(r\right)&={}_2F_1\left(\frac{1}{2}\left( 1- \frac{i\omega L^2}{r_h}- \frac{i J L}{r_h}-\nu\right),\frac{1}{2}\left(1-\frac{i \omega L^2}{r_h}- \frac{i J L}{r_h}+\nu \right);1-\frac{i J L}{r_h},{\frac{r^2}{r_h^2}}\right) \\
    H\left(r\right)&={}_2F_1\left(\frac{1}{2}\left( 1- \frac{i\omega L^2}{r_h}+ \frac{i J L}{r_h}-\nu\right),\frac{1}{2}\left(1-\frac{i \omega L^2}{r_h}+ \frac{i J L}{r_h}+\nu \right);1+ \frac{i J L}{r_h},{\frac{r^2}{r_h^2}}\right).
\end{align}

Near the AdS boundary ($r\rightarrow \infty$), the radial solution \eqref{sol1} becomes
\begin{align}\label{sol3}
 &   \phi_{bdry}(r)\approx -i r^{\frac{i\omega L^2}{r_h}-\nu-\frac{1}{2}} r_h^{1-\frac{i\omega L^2}{r_h}-\frac{i J L}{r_h}+\nu} (r^2-r_h^2)^{-\frac{i\omega L^2}{2 r_h}} e^{-\frac{\pi L (J+\omega L)}{2 r_h}}\times \nonumber \\
  & \times \Biggl( e^{-i\frac{\pi}{2}\nu}\Big(   \gamma\left(J, -\nu \right) C_1+ \gamma\left(-J, -\nu \right) C_2 \Big)+O\left(1/r^{3/2}\right) 
    +\frac{r^{2\nu}e^{i\frac{\pi}{2}\nu}}{r_h^{2\nu}}\left(  \gamma\left(J, \nu\right) C_1
    + \gamma\left(-J, \nu\right) C_2   +O\left(1/r^{3/2}\right) \right)\Biggr),
\end{align}
where
\begin{align}
    \gamma\left(J, \nu \right) &\equiv \frac{\Gamma(1-\frac{i J L}{r_h})\Gamma(\nu)}{\Gamma\left(\frac{1}{2}(1-\frac{i \omega L^2}{r_h}-\frac{i J L}{r_h}+\nu)\right) \Gamma\left(\frac{1}{2}(1+\frac{i \omega L^2}{r_h}-\frac{i J L}{r_h}+\nu)\right) }, 
      \end{align}
Normalizability at $r\rightarrow \infty$ sets the 2nd term of equation \eqref{sol3} to zero, which leads to
\begin{equation}\label{c1c2}
    C_2=-\frac{\gamma\left(J,\nu\right)}{\gamma\left(-J,\nu \right)}C_1,
\end{equation}
fixing the constant of integration $C_2$ in terms of $C_1$ or vice versa. 


We will eventually place our boundary condition at a stretched horizon, to be thought of as a Planck length or so outside the horizon. Near the horizon, the radial solution can be approximated as 
\begin{equation}\label{sol5}
    \phi_{hor}(r)\approx C_{1} \left(P_1\left(r/r_h-1\right)^{-\frac{i \omega L^2}{2 r_h}}+Q_1 \left(r/r_h-1 \right)^{\frac{i\omega L^2}{2r_h}} \right),
\end{equation}
where
\begin{align}
        P_1 &=-\frac{2^{-\frac{i\omega L^2}{2r_h}}e^{-\frac{\pi J L}{r_h}}\left(J \pi L\right) \left(e^{\frac{2\pi J L}{r_h}}-1 \right) r_h^{-\frac{1}{2}-\frac{i\omega L^2}{ r_h}-\frac{i J L}{r_h}} \text{csch}(\frac{\pi\omega L^2}{r_h})\Gamma(-\frac{i J L}{r_h})}{
    \left(e^{\frac{\pi J L}{r_h}}+e^{\pi(i\nu+\frac{\omega L^2}{r_h})}\right)\Gamma(1-\frac{i\omega L^2}{r_h})\Gamma(\frac{1}{2}(1+ \frac{i\omega L^2}{r_h}-\frac{i J L}{r_h}-\nu))
    \Gamma(\frac{1}{2}(1+ \frac{i\omega L^2}{r_h}-\frac{i J L}{r_h}+\nu))} \label{PQ1}\\
    Q_1 &=\frac{(-1)^{\frac{i\omega L^2}{r_h}} 2^{1+\frac{i\omega L^2}{2 r_h}}e^{\frac{2\pi\omega L^2}{r_h}}\pi^2  r_h^{\frac{1}{2}-\frac{i\omega L^2}{ r_h}-\frac{i J L}{r_h}} (\text{coth}(\frac{\pi \omega L^2}{r_h})-1)}{
    \left(e^{i\pi\nu}+e^{\frac{\pi L(J+\omega L)}{r_h}}\right)\Gamma(\frac{i J L}{r_h})\Gamma(1+\frac{i\omega L^2}{r_h})\Gamma(\frac{1}{2}(1-\frac{i\omega L^2}{r_h}-\frac{i J L}{r_h}-\nu))\Gamma(\frac{1}{2}(1-\frac{i\omega L^2}{r_h}-\frac{i J L}{r_h}+\nu))}.\label{PQ2}
\end{align}

In \cite{synthfuzz} we demanded a vanishing condition for the scalar at the stretched horizon $r=r_0$. Motivated by the angle-dependent profiles that are found in BPS fuzzballs, we will generalize this in the present paper. We will demand that at $r=r_0$ the scalar field takes the form of a given profile $\phi_0(\psi)$. In terms of the notation introduced in \eqref{scalarmodes}, we will write
\begin{eqnarray}
\frac{1}{\sqrt{r_0}} \sum_{J,\omega}   e^{i J\psi}e^{-i \omega t} \phi_{\omega, J}(r_0) = \phi_{0}(\psi,t) \label{profile-def}
\end{eqnarray}
Expanding the RHS in terms of the Fourier modes $e^{i J\psi}$ and $e^{-i\omega t}$  and absorbing some constants suitably, we get an equation of the form 
$\phi_{hor}(r=r_0)=C_0$ where on both LHS and RHS we have suppressed the $\omega$ and $J$ subscripts. Note that ultimately we will get a quantization condition for our $\omega$'s, and this means that an implicit assumption in the above approach is that the $\phi_0(\psi,t)$ can be expanded in terms of these modes. Our explicit boundary conditions below and their solution can be viewed as a self-consistent way to do this. 

Concretely, this leads to 
\begin{align}
    & C_1 \left( P_1 \left( r_0/r_h-1\right)^{-\frac{i\omega L^2}{2r_h}}+ Q_1 \left( r_0/r_h-1 \right)^{\frac{i\omega L^2}{2r_h}}\right)= C_0,\\
       & \implies \frac{P_1}{Q_1} = \frac{C_0 }{C_1 Q_1}\left( r_0/r_h-1\right)^{\frac{i\omega L^2}{2 r_h}}-\left( r_0/r_h-1\right)^{\frac{i\omega L^2}{r_h}}.\label{keyeqBTZ}
\end{align}
As in \cite{synthfuzz}, it is possible to show that $|P_1|=|Q_1|$.  So by writing  $P_1 = |P_1| e^{i\alpha}$  and $Q_1= |Q_1| e^{i\beta}$, \eqref{keyeqBTZ} can be written as
\begin{eqnarray}
e^{i(\alpha-\beta)} &=& \mu_J e^{i\left(\frac{\lambda_J\omega L^2 }{r_h}+\frac{\theta}{2}\right)}-e^{i\theta}\label{geneq2}
\end{eqnarray}
where
\begin{eqnarray}
\theta=\text{Arg}\left[\left( r_0/r_h-1\right)^{\frac{i\omega L^2}{r_h}}\right], \ \ \mu_J=\Big|\frac{C_{0} }{C_{1} Q_1}\Big|, \ \ {\rm and} \ \ \frac{\lambda_J \omega L^2}{r_h}=\text{ Arg}\left[\frac{C_{0} }{C_{1} Q_1}\right] \label{threeeq}
\end{eqnarray}
We have emphasized the $J$-dependence of $\mu$ and $\lambda$ in the notation, but it should be noted that with these definitions, they have an $n$-dependence as well. 
The real and imaginary parts of \eqref{geneq2} lead to the definition
\begin{eqnarray}
\mu_J = 2\cos\left(\frac{ \lambda_J \omega L^2 }{r_h}-\frac{\theta}{2}\right) \label{mueqBTZ}
\end{eqnarray}
as well as the quantization condition on $\omega$,
\begin{eqnarray}
 \cos(\alpha-\beta)=\cos\left(\frac{2\lambda_J \omega L^2}{r_h}\right), \ \ \sin(\alpha-\beta)=\sin\left(\frac{2\lambda_J \omega L^2}{r_h}\right). \label{lambdaeqBTZ}
\end{eqnarray}
These last equations are key equations for our purposes. Since this is a phase equation, the modes depend on a free integer $n$. It is possible to check that these two equations together reduce to the quantization condition we had in \cite{synthfuzz} when we set $\mu_J =0$. More generally, one can solve the quantization condition by choosing $\lambda_J$ from a distribution, which we will usually take to be Gaussian. 

We will take $\lambda$ for each value of $J$ from the same distribution. Note that heuristically, $\lambda_J$ is comparable to the stretched horizon location. One way to see this is to note that \eqref{mueqBTZ} implies (if there are no fluctuations, and $\lambda$ and $\mu$ are taken to be $J$-independent constants) that fixing
\begin{eqnarray}
\lambda_J = \frac{1}{2}\ln \left(\frac{r_0}{r_h} -1\right) \label{mean}
\end{eqnarray}
fixes $\mu_J$. More generally, the fact that the difference between $\lambda_J$  and $\frac{1}{2}\ln \left(\frac{r_0}{r_h} -1\right)$ is what shows up in \eqref{mueqBTZ} suggests that the natural scale of $\lambda_J$ is the stretched horizon radius in (essentially) tortoise coordinates. Eqn \eqref{mueqBTZ} also makes it tempting to view the fluctuations in $\mu_J$ as due {\em not} to the fluctuations in $\lambda_J$ but due to the fluctuations of the stretched horizon. This last interpretation is of course simply a heuristic, because it is not meaningful to have a $J$-dependent notion of stretched horizon radius. Nonetheless, we view this as highly suggestive, in light of the usual claim that the profile functions in fuzzball geometries are supposed to capture the fluctuations of the cap. Indeed, our initial motivation for considering the scalar field profile, was as a proxy for this. 

It is worth emphasizing in the above discussion (and elsewhere), that there is some leftover freedom in fixing $C_1$ in terms of $C_0$ and the rest of the quantities. An analogous freedom existed in  \cite{synthfuzz} as well -- our demands do not completely fix the boundary conditions, but they fix them enough to determine the normal modes. We can fix this extra freedom by setting $C_1Q_1=1$ so that $\mu_J$ and $\lambda_J$ have the nice interpretation as (essentially) the magnitude and phase of $C_0$. Remember that $C_0$ has $J$-dependence which we often suppress to avoid notational congestion, it is the Fourier coefficient of the scalar profile.  

There is one choice we have made in the above definitions, which may be worth further study. In defining $\lambda_J$ via the last equation in \eqref{threeeq}, we have extracted an $\omega$ on the LHS. It may  also be natural to define the $\lambda$ variable without this, in which case our quantization conditions should be solved after the replacement $\lambda_J \rightarrow \lambda_J/\omega$ and choosing the new $\lambda$'s from some suitable distribution.  Since the target results we are aiming for are believed to be robust semi-qualitative statements like level repulsion and the linear ramp, these choices should not affect them. We have checked that indeed this is the case. Ultimately these choices all correspond to how we parametrize the Fourier modes $C_0$ of the profile $\phi_0(\psi,t)$ in \eqref{profile-def}. Explicitly, the profile should be written as 
\begin{eqnarray}
 \phi_0(\psi,t)= \sum_{n,J} C_{0 (n,J)} e^{iJ \psi}e^{-i \omega(n,J) t} 
\end{eqnarray}
and our choice corresponds to the parametrization
\begin{eqnarray}
C_{0 (n,J)}=\mu_{J,n}e^{ i\frac{\lambda_{J}\omega(n,J) L^2 }{r_h}} \label{defC0}
\end{eqnarray}
where we have kept the $n$ and $J$ dependencies, fully explicit. If we absorb the $\omega$ into the definition of $\lambda$ as discussed above, then the $\mu$ (and therefore the $C_0$) have only $J$-dependence. (Superficially, this may seem illegal because $\omega$'s have an $n$-dependence. But remember that the $\omega$'s are determined {\em after} the definition of $\lambda$, so one can check that this is perfectly well-defined.) This leads to some nice features in some expressions, but also some compensating complications/ugliness in others. So we will stick to the form defined by \eqref{geneq2} and \eqref{threeeq}, or \eqref{defC0}. It may be interesting to investigate the naturalness of the choices involved here from the perspective of Haar typicality in the phase space of the scalar field, but we will not undertake it here.

With these caveats, one way to get some intuition for the profile is to consider the quantity 
\begin{eqnarray}
\tilde \phi(\psi)\equiv \sum_{J=0}^{J_{\rm cut}} C_{0 (n=0,J)} e^{iJ \psi} = \sum_{J=0}^{J_{\rm cut}}\mu_{J,n=0} e^{ i\frac{\lambda_{J}\omega(n=0,J) L^2 }{r_h}}  e^{iJ \psi}. \label{profile}
\end{eqnarray}
This is what we will often call the profile function. It should be emphasized that our quantization condition arises essentially as a condition on the {\em phase} of the Fourier coefficient. The various arbitrary choices we discussed above can be understood as arising from the fact that it does not unambiguously fix $C_0$. In writing the second equality of \eqref{profile} we have fixed $C_1 Q_1 =1$ as mentioned above, but this is an {\em ad-hoc} choice. Similar statements were true in the discussion in \cite{synthfuzz} as well, where the magnitude information was again not needed to determine the normal modes. One way to understand this in the present setting is to note that the last two equations in \eqref{threeeq} basically determine the phase and the magnitude of the profile $C_0$ via
\begin{eqnarray}
\mu_J e^{i\frac{\lambda_{J}\omega L^2 }{r_h}} = \frac{C_0}{C_1 Q_1}. \label{ratio}
\end{eqnarray}
Once we make a choice of $\lambda$ (which is a single real variable that captures the phase information) the quantization condition is obtained via \eqref{lambdaeqBTZ}. Then $\mu_J$ is completely fixed via \eqref{mueqBTZ}. All of this only fixes the ratio on the RHS of \eqref{ratio}, while the profile itself is controlled by $C_0$. Fourier series where the {\em phase} is suitably  random have been studied extensively by mathematicians, see eg. the book  \cite{RandomFourierBook}. It seems significant that this structure naturally arises in our discussions; this is clearly worthy of further study.


In the plots in this section, we have set $L=r_h=1$, and $\langle \lambda \rangle=\frac{1}{2}\ln \left(\frac{r_0}{r_h} -1\right)$, as we change the variance of the Gaussian distribution from which $\lambda$ is chosen. This choice of $\langle \lambda \rangle$ ensures that $\mu_J=2$  in the zero-variance limit. This is slightly different from the $\mu_J=0$ condition in \cite{synthfuzz} but it is natural (and straightforward to check) that the qualitative results on LSD and SFF remain identical. One can also in principle treat $\mu_J$ (instead of $\lambda_J$) as the quantity chosen from a distribution. This is slightly more convenient to connect to the language of \cite{synthfuzz}. This changes some of our formulas in minor ways, but the essential point that there is {\em one} real parameter worth of freedom that we are fixing, remains intact. 
We have experimented  with various choices of $\lambda$-variance as a function of $J$, eg.  $\sigma_{\lambda_J} \equiv \sigma_0, \sigma_0/J, \sigma_0/\sqrt{J}$. In the plots in this section, we present the $\sigma_0/\sqrt{J}$ case and we quote the value of $\sigma_0$. We will sometimes refer to $\sigma_0$ loosely as the variance. A suppression of the variance with $J$ is useful because the normal mode level-spacing gets smaller as $J$ increases, and therefore too large a variance can completely destabilize the structure of the spectrum (and along with it, the linear ramp and level repulsion). Let us also mention that when we juxtapose the plots of an SFF and an LSD for some choice of variance, we show it for the {\em same} realization that we choose from the Gaussian distribution. This statement applies to the Rindler plots of the next section as well.

For zero variance, we reproduce the ``extreme'' Wigner-Dyson plots for the level spacing that we found in \cite{synthfuzz} as well as the linear ramp. If we increase the variance slightly, the ramp remains intact, but the level-spacing takes the more conventional WD form. We can find fits with GSE, GUE or GOE with minor increments in variance, we present GUE in the plots. Eventually, as we increase the variance to very large values, the level spacing degenerates to a Poisson form and the ramp is lost. 

\begin{figure}[H]
\begin{subfigure}{0.47\textwidth}
    \centering
    \includegraphics[width=\textwidth]{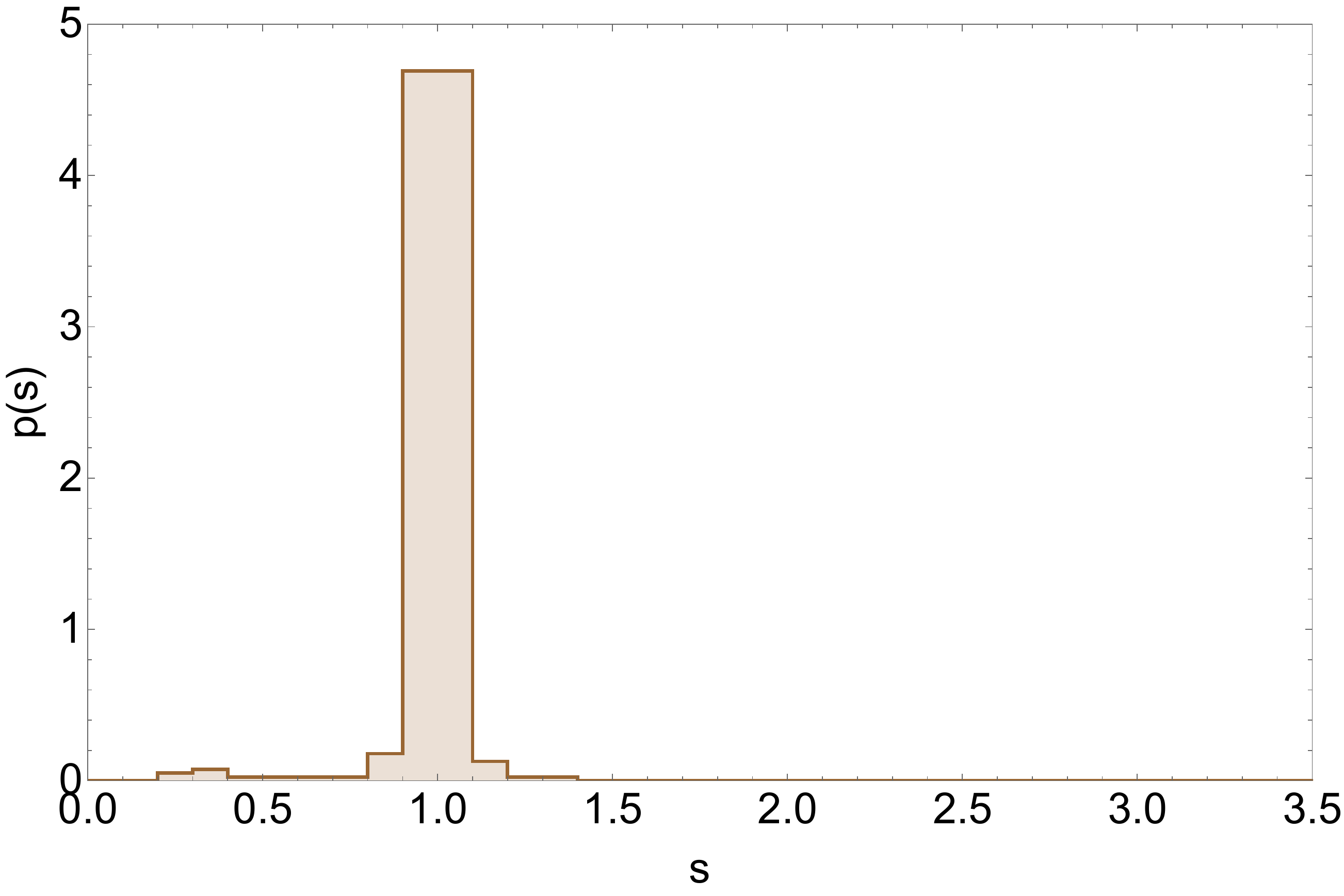}
    \end{subfigure}
    \hfill
    \begin{subfigure}{0.47\textwidth}
    \includegraphics[width=\textwidth]{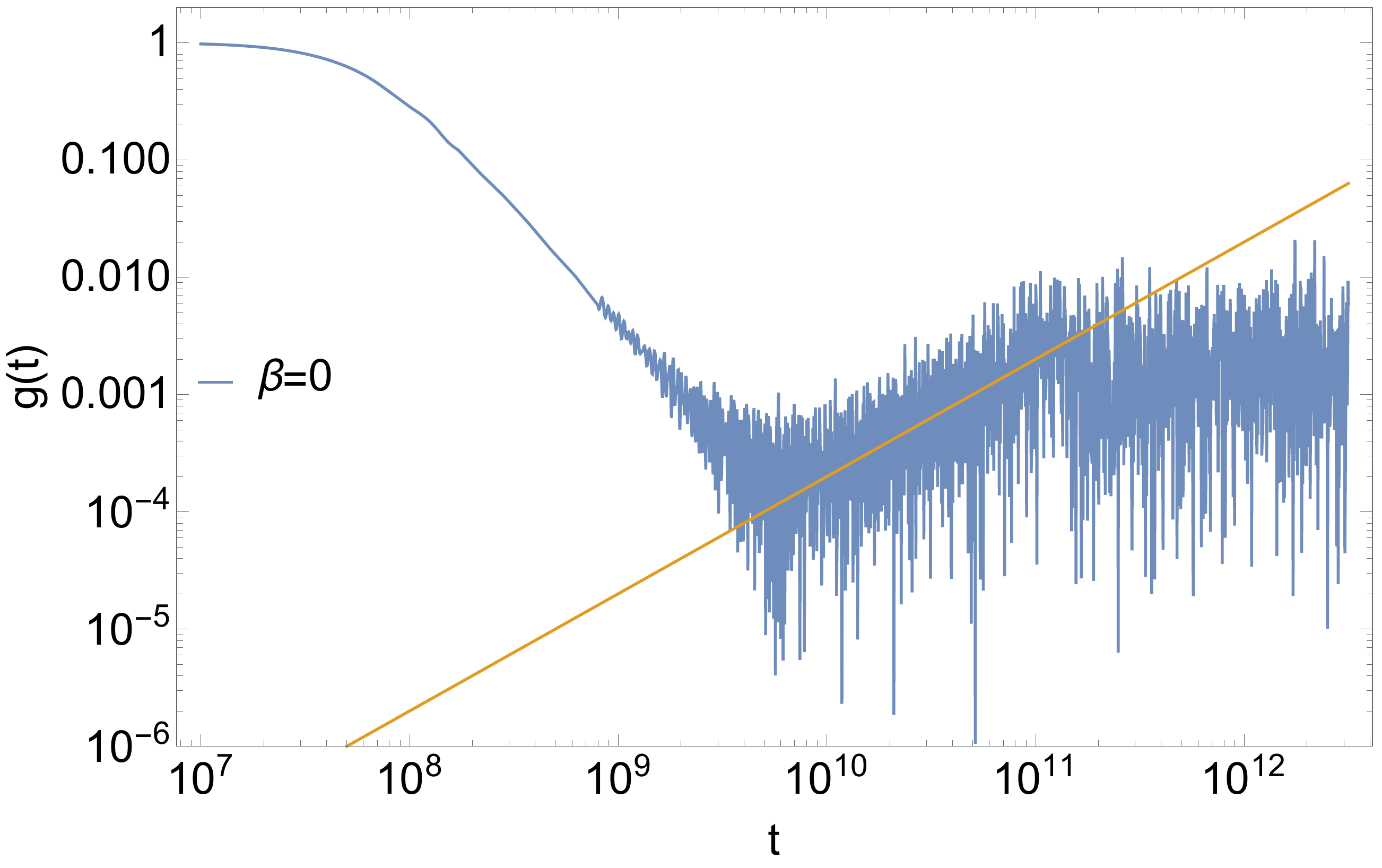}
    \end{subfigure}
    \caption{LSD (left) and SFF (right) for BTZ with $\langle \lambda \rangle = -10^4$ and $J_{max}=400$ with $\sigma_0=0.0$. We are working with $\omega(n=2,J)$. These results are a version of the results in \cite{synthfuzz}.} 
    \label{BTZ_extreme}
\end{figure}

\begin{figure}[H]
\begin{subfigure}{0.47\textwidth}
    \centering
    \includegraphics[width=\textwidth]{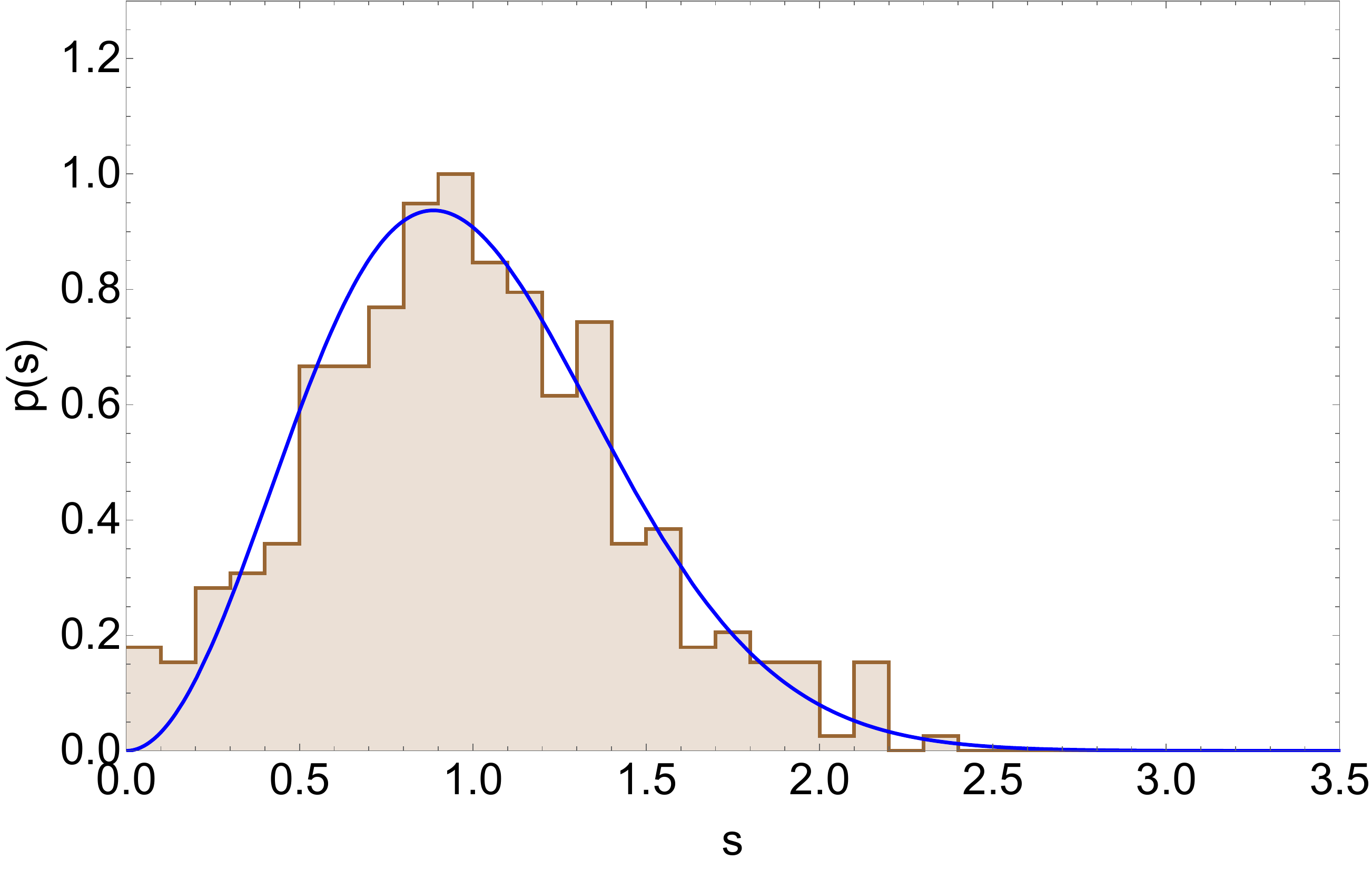}
    \end{subfigure}
    \hfill
    \begin{subfigure}{0.47\textwidth}
    \includegraphics[width=\textwidth]{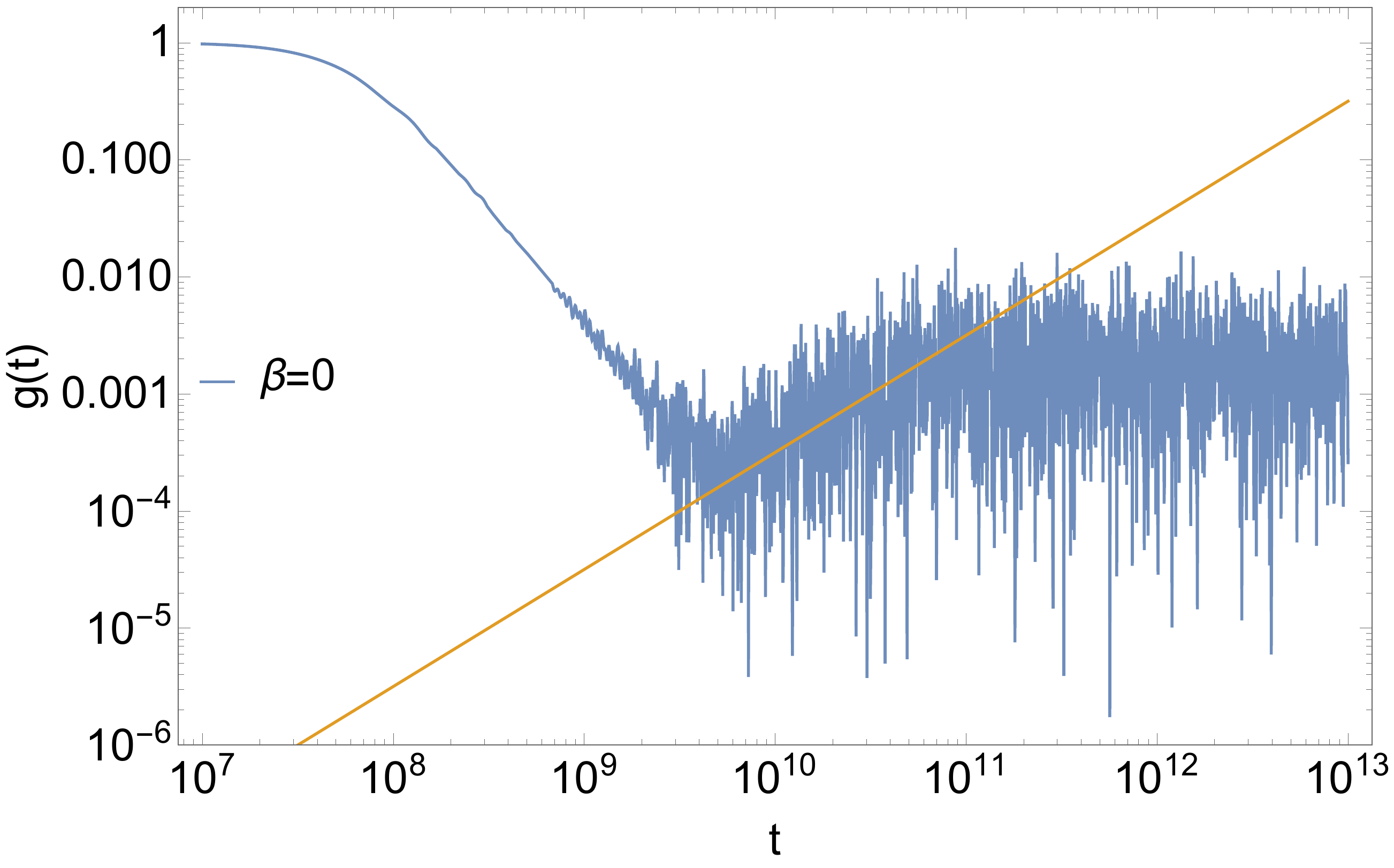}
    \end{subfigure}
    \caption{LSD (left) and SFF (right) for BTZ with $\langle \lambda \rangle = -10^4$ and $J_{max}=400$ with $\sigma_0=0.025$. We are working with $\omega(n=2,J)$. The blue curve on the left is GUE. } 
    \label{BTZ_GUE}
\end{figure}

\begin{figure}[H]
\begin{subfigure}{0.47\textwidth}
    \centering
    \includegraphics[width=\textwidth]{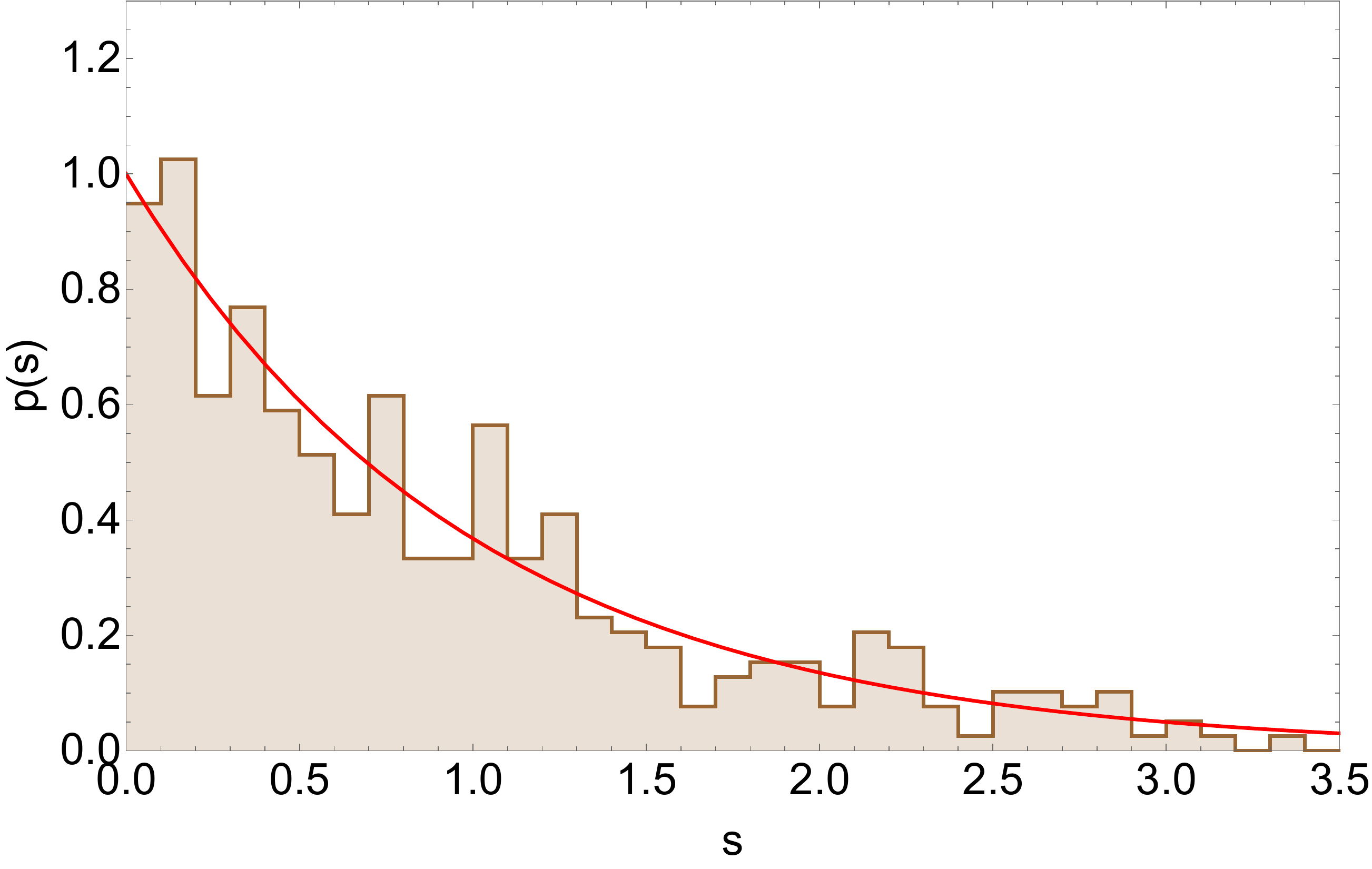}
    \end{subfigure}
    \hfill
    \begin{subfigure}{0.47\textwidth}
    \includegraphics[width=\textwidth]{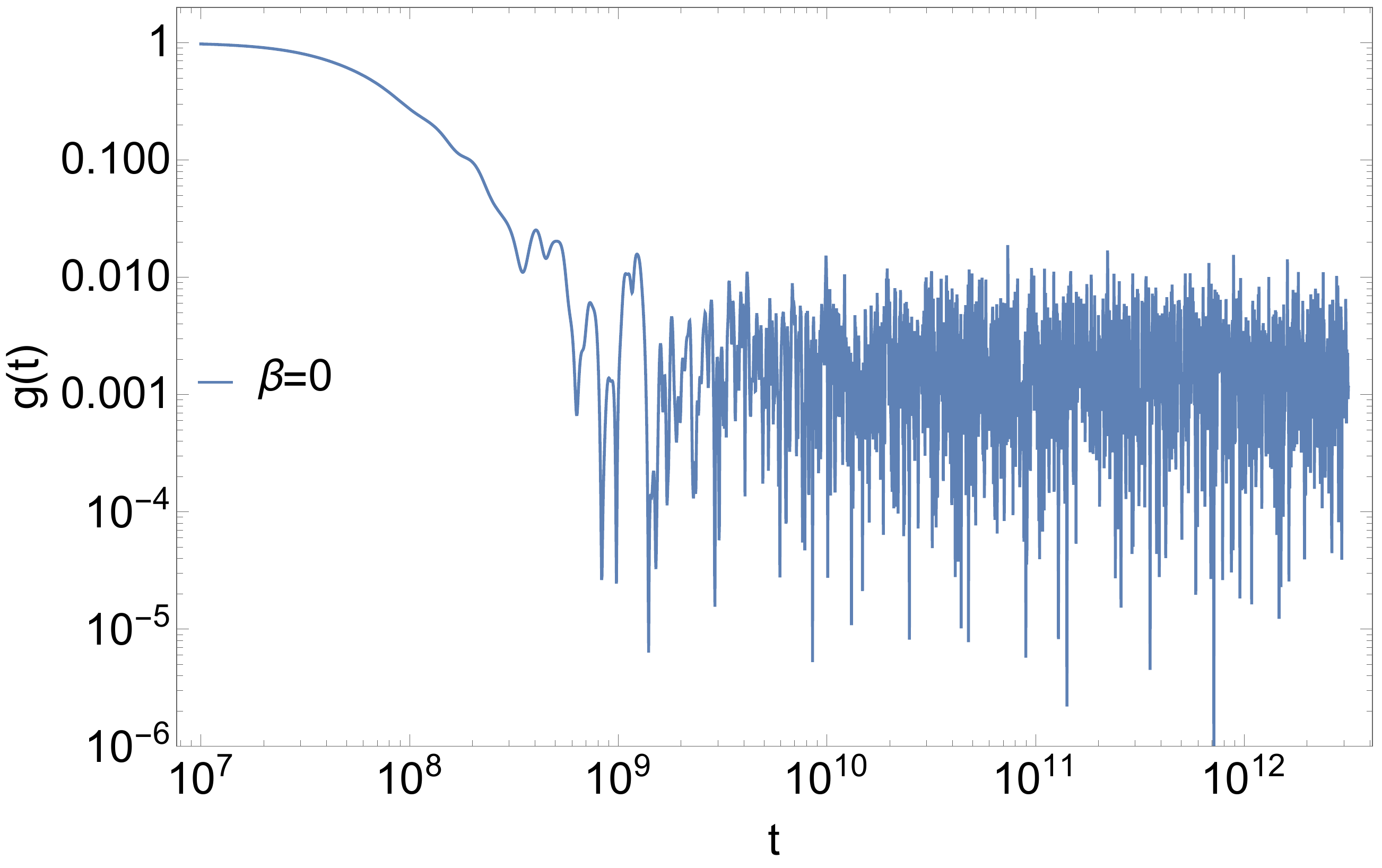}
    \end{subfigure}
    \caption{LSD (left) and SFF (right) for BTZ with $\langle \lambda \rangle = -10^4$ and $J_{max}=400$ with $\sigma_0=2.0$. We are working with $\omega(n=2,J)$. The red curve on the left is Poisson. } 
    \label{BTZ_Poisson}
\end{figure}

\begin{figure}[H]
\begin{subfigure}{0.47\textwidth}
    \centering
    \includegraphics[width=\textwidth]{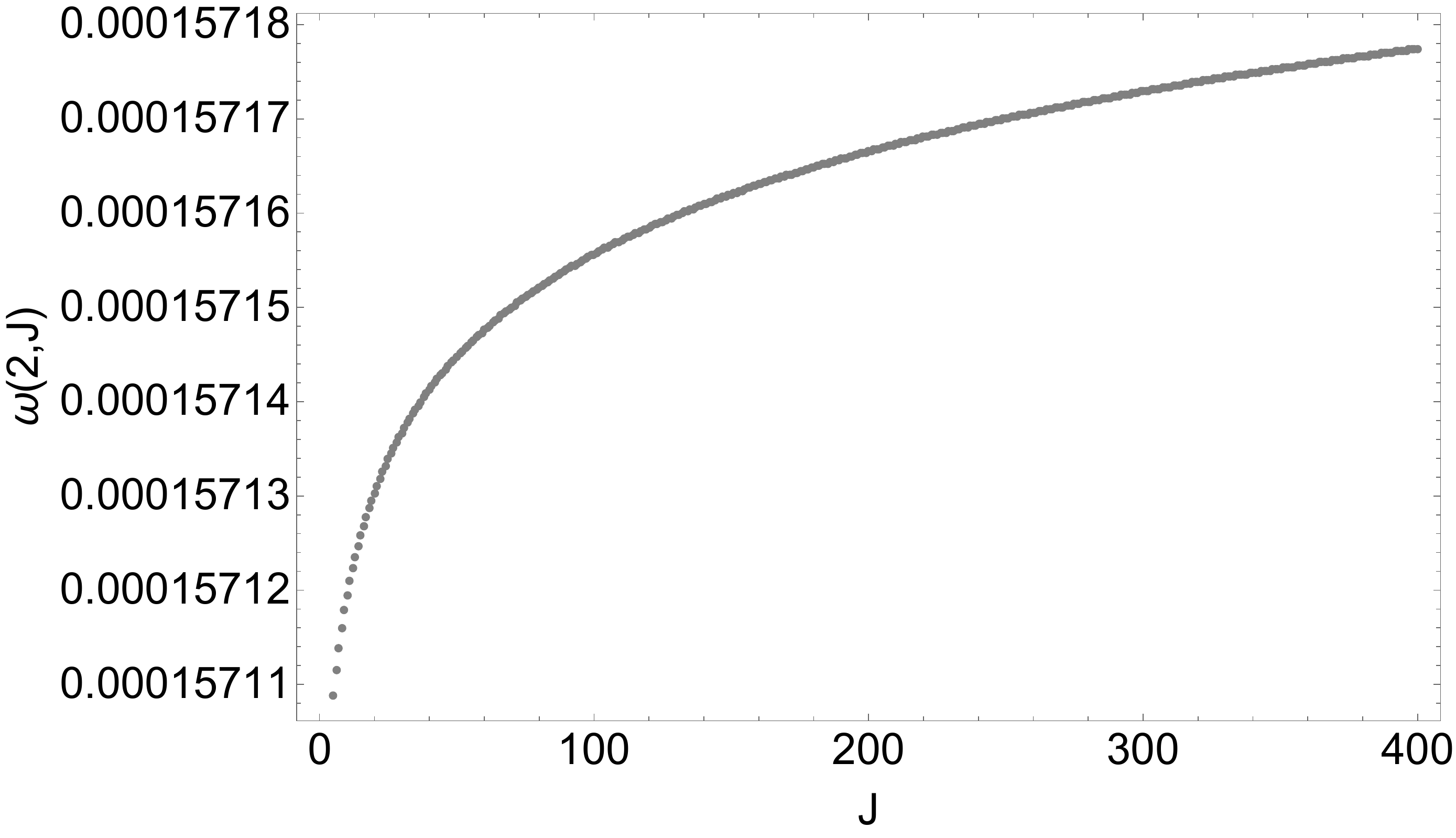}
    \end{subfigure}
    \hfill
    \begin{subfigure}{0.47\textwidth}
    \includegraphics[width=\textwidth]{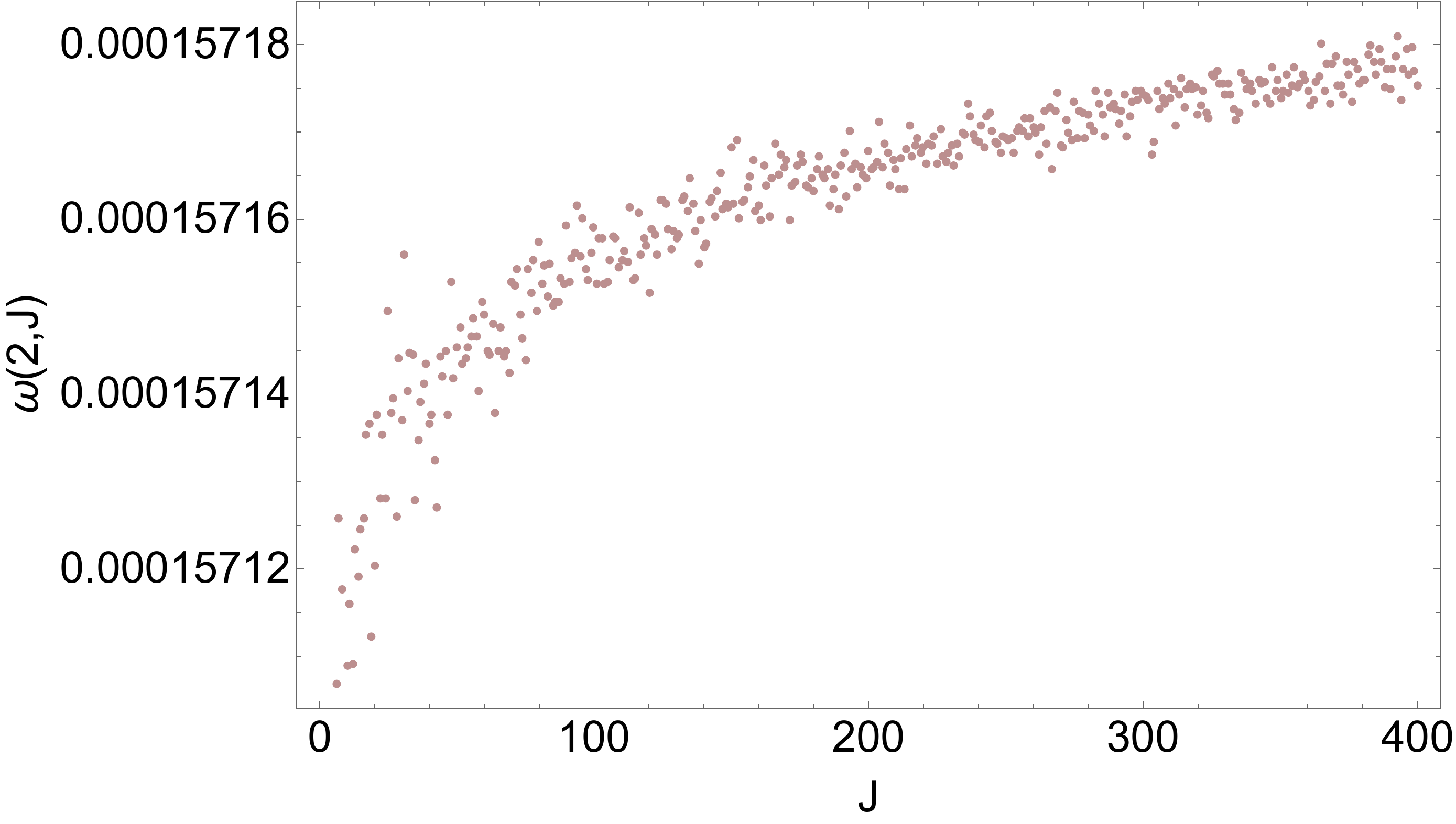}
    \end{subfigure}
    \caption{Spectrum of BTZ with $\sigma_0=0$ (left) vs $\sigma_0=2.0$ (right). $\langle \lambda \rangle = -10^4$ and $J_{max}=400$. We show $\omega(n=2,J)$. } 
    \label{BTZ_spectra}
\end{figure}


\section{Case Study: Rindler $\times$ Compact Space}

We will follow the motivations and discussion in section 4.2 of \cite{synthfuzz} when developing the Rindler case, which the reader should consult for notations. We solve the wave equation in the metric  
\begin{equation}
    ds^2=e^{2 a \xi}(-d\eta^2+d\xi^2)+R^2d\phi^2
\end{equation}
%
and introduce $A\equiv \omega/a$ and $y\equiv e^{a \xi}(J/a R)$ as in \cite{synthfuzz}. In terms of $y$ variable the position of boundary and horizon are given by $y\rightarrow\infty$ and $y\rightarrow0$ respectively. In the notations of \cite{synthfuzz}, we require that the field $\phi(y)$ vanish at boundary.  We also demand that it has a profile at some small $y_0$ (or $\xi=\xi_0$). 
When $y\rightarrow\infty$, the relevant equation is \cite{synthfuzz} 
\begin{eqnarray}
    \phi(y)\rightarrow(C_1+C_2)\frac{e^y}{\sqrt{2\pi y}}+(C_1 e^{\pi A}+C_2 e^{-\pi A})\frac{e^{-y}}{\sqrt{2\pi y}}.
\end{eqnarray}
The boundary condition at infinity leads to $C_1=-C_2$, and at $y_0$ implies (in notation that is parallel to the BTZ case before):
\begin{align}
    &C_1(I[-i A,y_0]-I[i A,y_0])=C_0, \nonumber\\
     &\implies I[-i A,y_0]-I[i A,y_0]=\frac{C_{0}}{C_1}
\end{align}
Near horizon i.e. in the limit $y_0\rightarrow0$ the above expressions can be approximated by
\begin{eqnarray}
C_1 \left( y^{-iA}\frac{2^{i A}}{\Gamma(1-i A)}-y^{i A}\frac{2^{-i A}}{\Gamma(1+i A)}\right) &=& C_{0}\\
-\frac{C_{0}}{C_1}\left( \frac{J}{a R}\right)^{-i A}\left(\frac{e^{a\xi}}{2} \right)^{i A}\Gamma(1+i A)-\left(\frac{e^{a\xi}}{2}\right)^{2 i A}
&=& \left( \frac{J}{aR}\right)^{-2i A}\frac{\Gamma(iA)}{\Gamma(-iA)}\label{rind4}
\end{eqnarray}
Now ${\rm Abs}\left[ \left( \frac{J}{aR}\right)^{-2i A}\frac{\Gamma(iA)}{\Gamma(-iA)} \right]=1$, so \eqref{rind4} can be written, again in notation that simulates the BTZ case as
\begin{eqnarray}
\mu_J e^{i\omega\lambda_J} e^{i\theta/2}-e^{i\theta} &=& e^{i\alpha}\label{rind5}
\end{eqnarray}
with 
\begin{eqnarray}
&\mu_J = {\rm Abs}\left[-\frac{C_0}{C_1}\left( \frac{J}{a R}\right)^{-i A}\Gamma(1+iA)\right],  \ \  \omega\lambda_J = {\rm Arg}\left[-\frac{C_0}{C_1}\left( \frac{J}{a R}\right)^{-i A}\Gamma(1+iA)\right], \ \ \nonumber \\
&\alpha = {\rm Arg}\left[\left( \frac{J}{aR}\right)^{-2i A}\frac{\Gamma(iA)}{\Gamma(-iA)}\right], \ \
\theta = {\rm Arg}\left[\left( \frac{e^{a\xi}}{2}\right)^{2iA}\right] \nonumber \\
\end{eqnarray}

\begin{figure}[H]
\begin{subfigure}{0.47\textwidth}
    \centering
    \includegraphics[width=\textwidth]{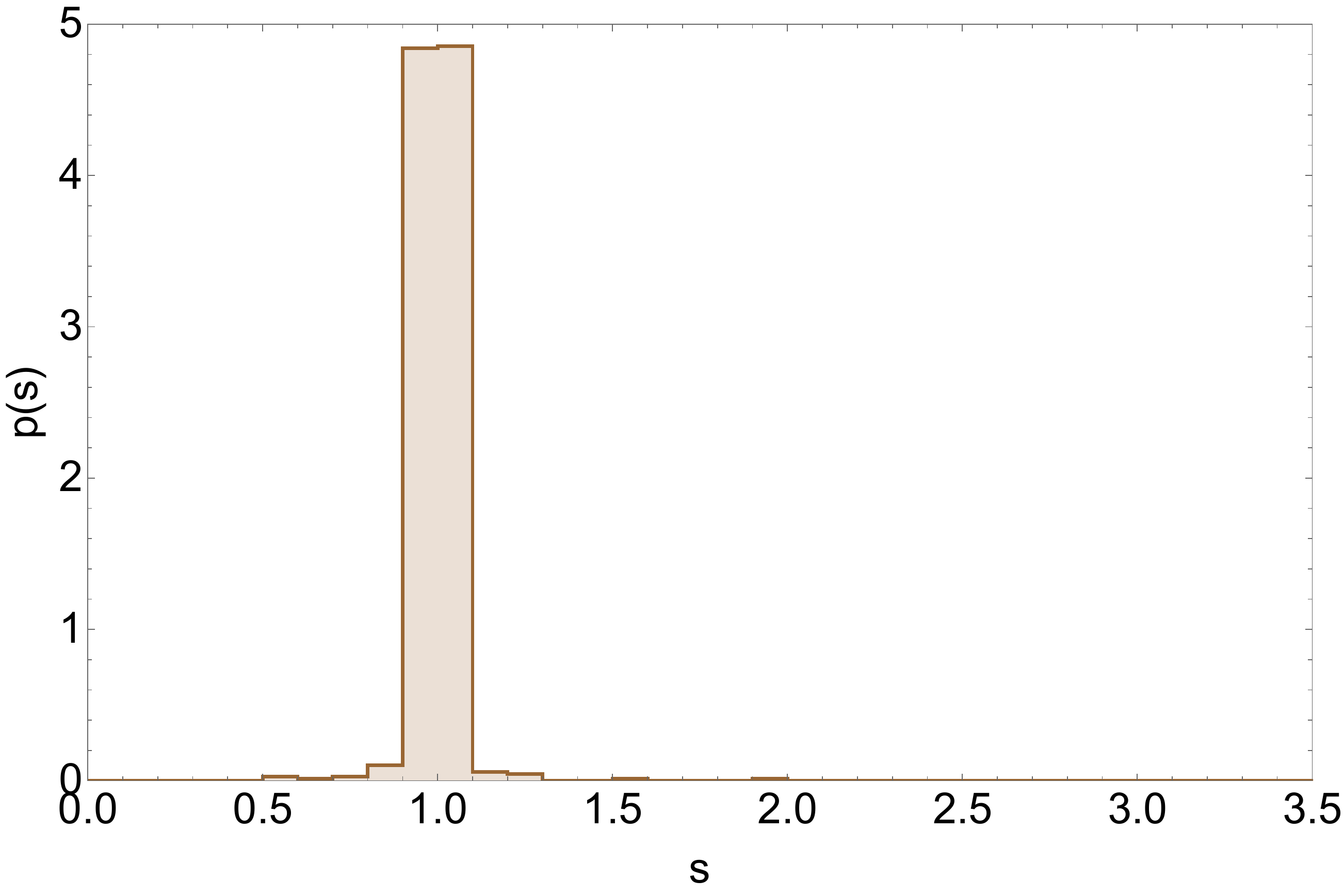}
    \end{subfigure}
    \hfill
    \begin{subfigure}{0.47\textwidth}
    \includegraphics[width=\textwidth]{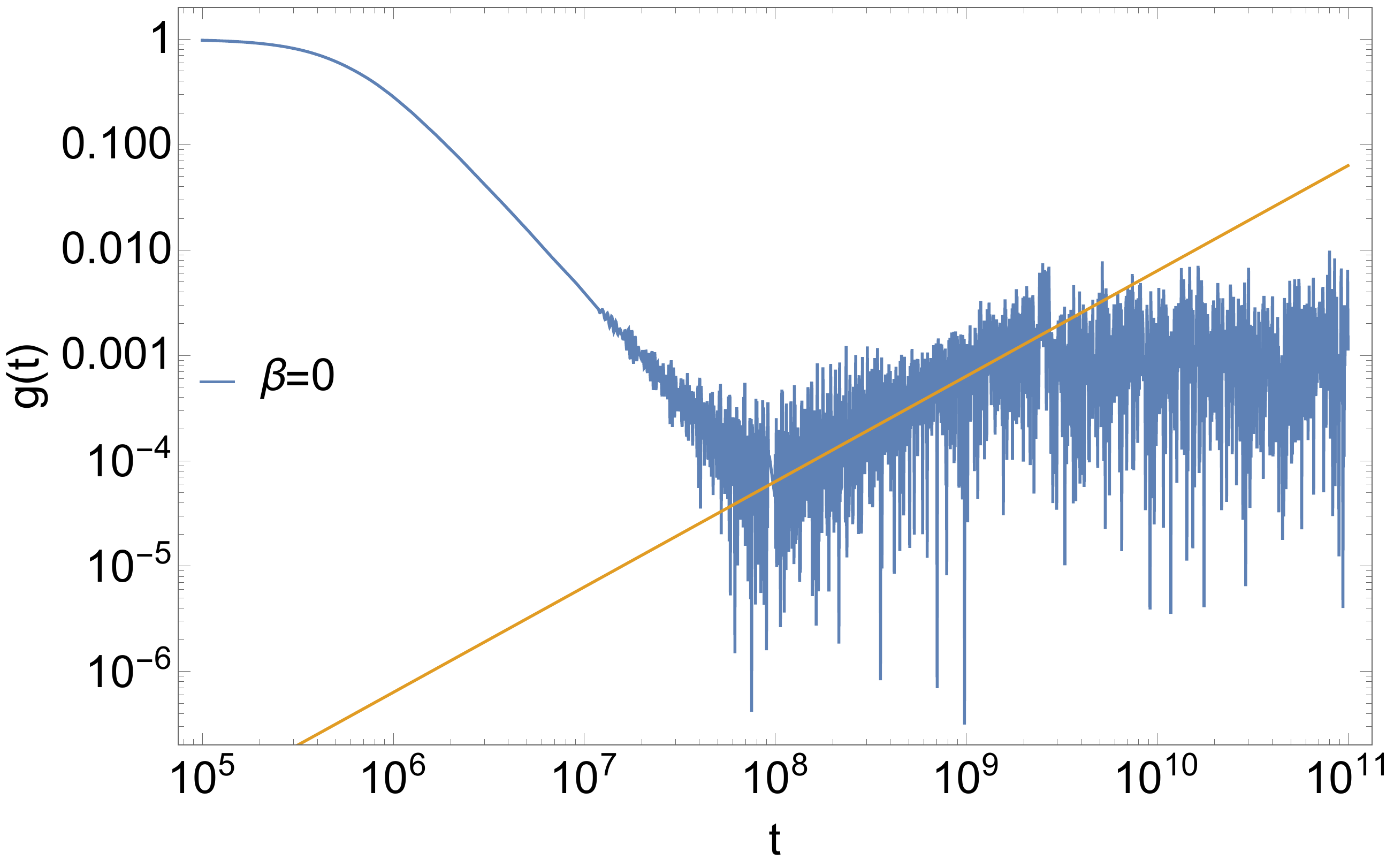}
    \end{subfigure}
    \caption{LSD (left) and SFF (right) for Rindler with parameters described in the text. $\sigma_0=0.0$. We are working with $\omega(n=1,J)$. These results should be compared to the results in \cite{synthfuzz}.} 
    \label{Rin_extreme}
\end{figure}

\begin{figure}[H]
\begin{subfigure}{0.47\textwidth}
    \centering
    \includegraphics[width=\textwidth]{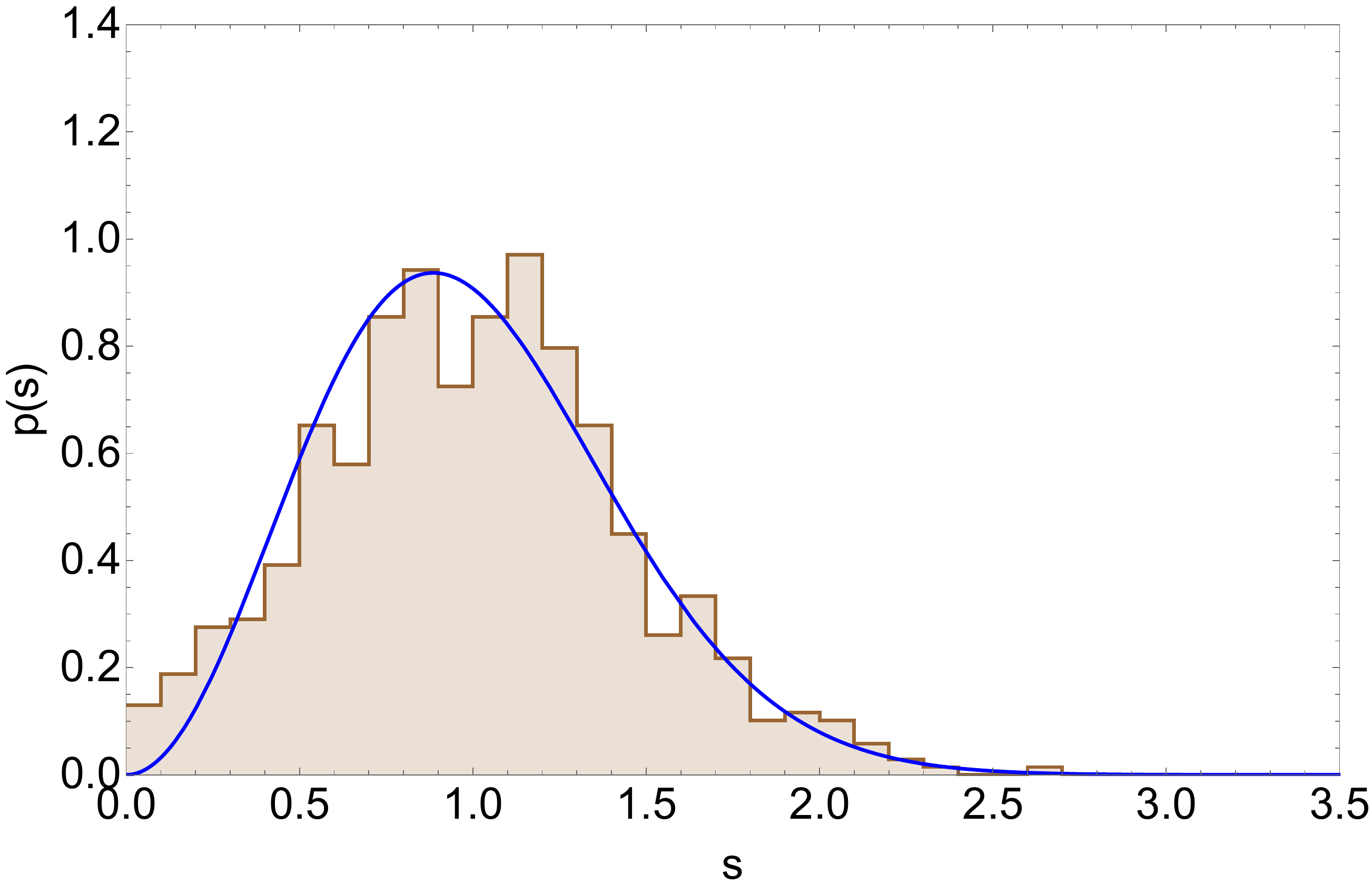}
    \end{subfigure}
    \hfill
    \begin{subfigure}{0.47\textwidth}
    \includegraphics[width=\textwidth]{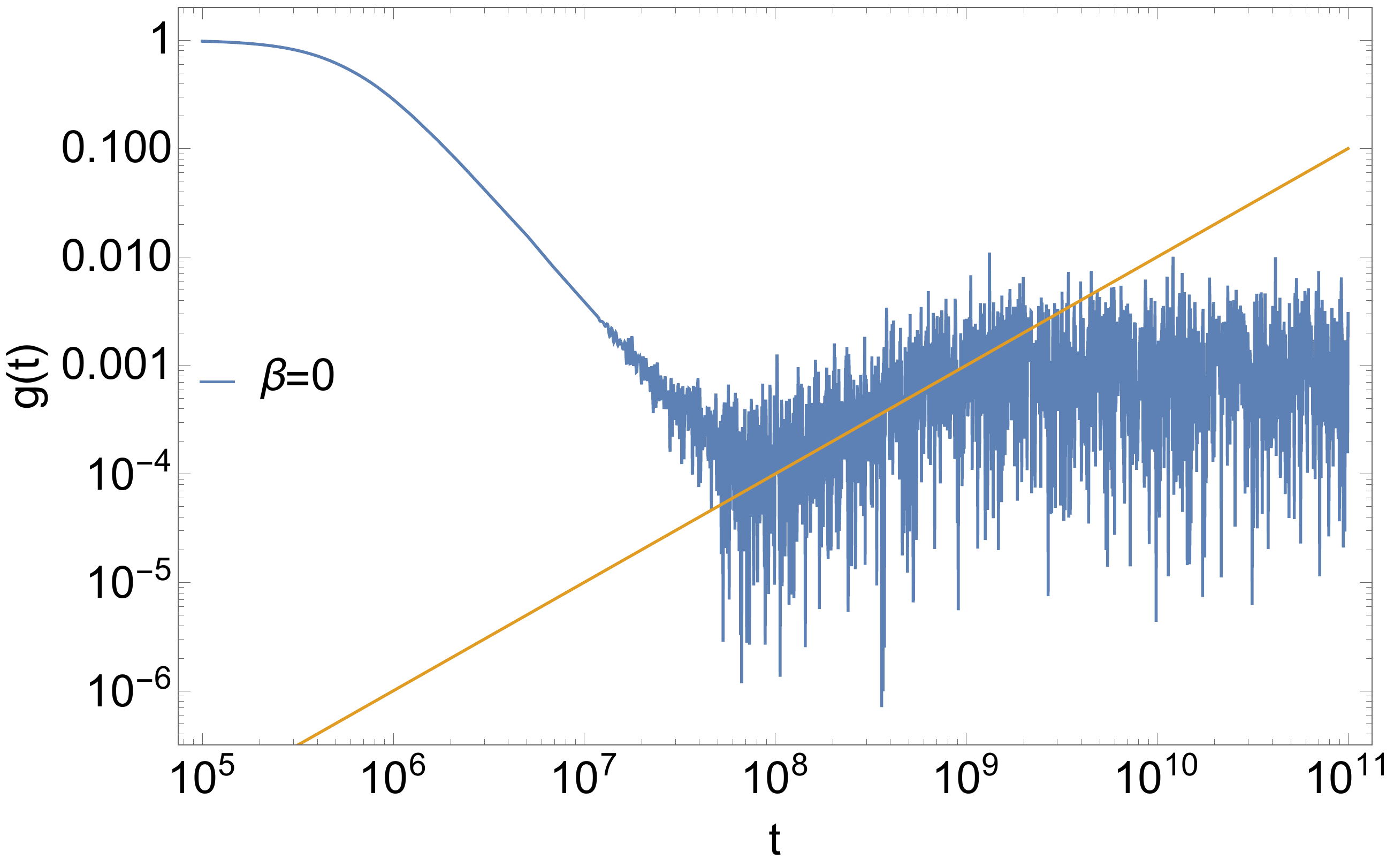}
    \end{subfigure}
    \caption{LSD (left) and SFF (right) for Rindler with parameters described in the text. $\sigma_0=0.019$. We are working with $\omega(n=1,J)$. The blue curve on the left is GUE.} 
    \label{Rin_GUE}
\end{figure}

\begin{figure}[H]
\begin{subfigure}{0.47\textwidth}
    \centering
    \includegraphics[width=\textwidth]{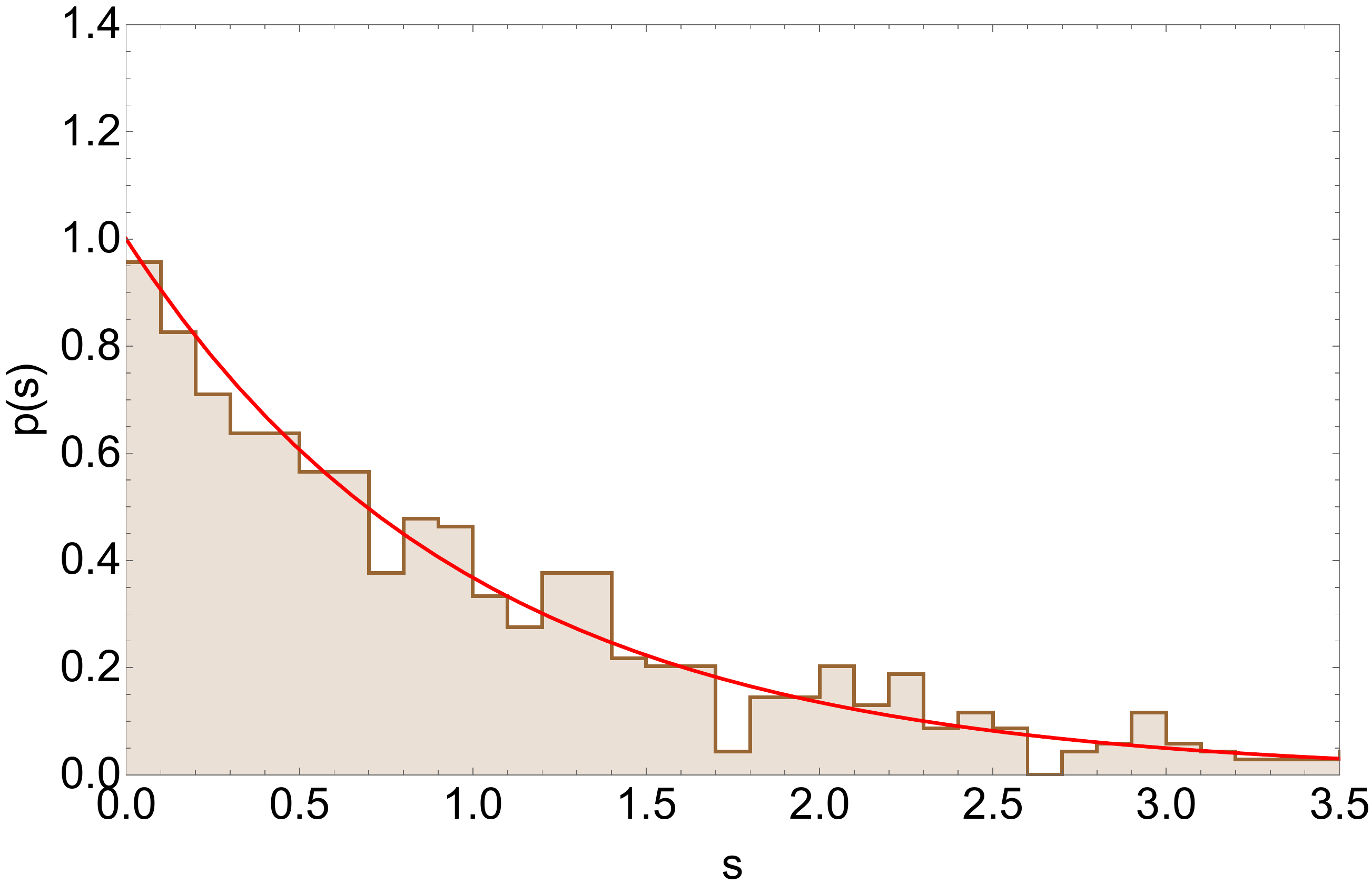}
    \end{subfigure}
    \hfill
    \begin{subfigure}{0.47\textwidth}
    \includegraphics[width=\textwidth]{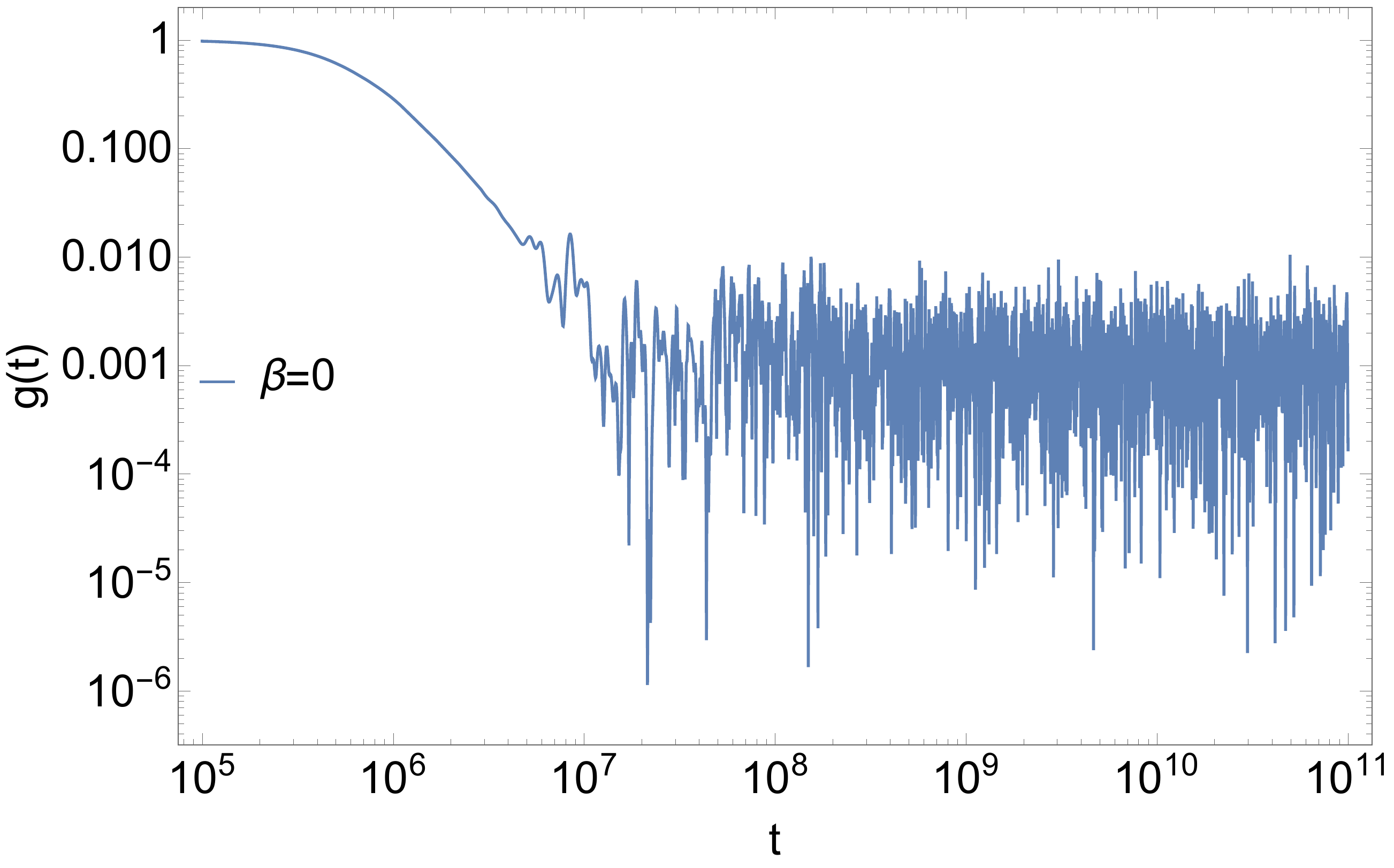}
    \end{subfigure}
    \caption{LSD (left) and SFF (right) for Rindler with parameters described in the text. $\sigma_0=1$. We are working with $\omega(n=1,J)$. The red curve on the left is Poisson.} 
    \label{Rin_Poisson}
\end{figure}

\begin{figure}[H]
\begin{subfigure}{0.47\textwidth}
    \centering
    \includegraphics[width=\textwidth]{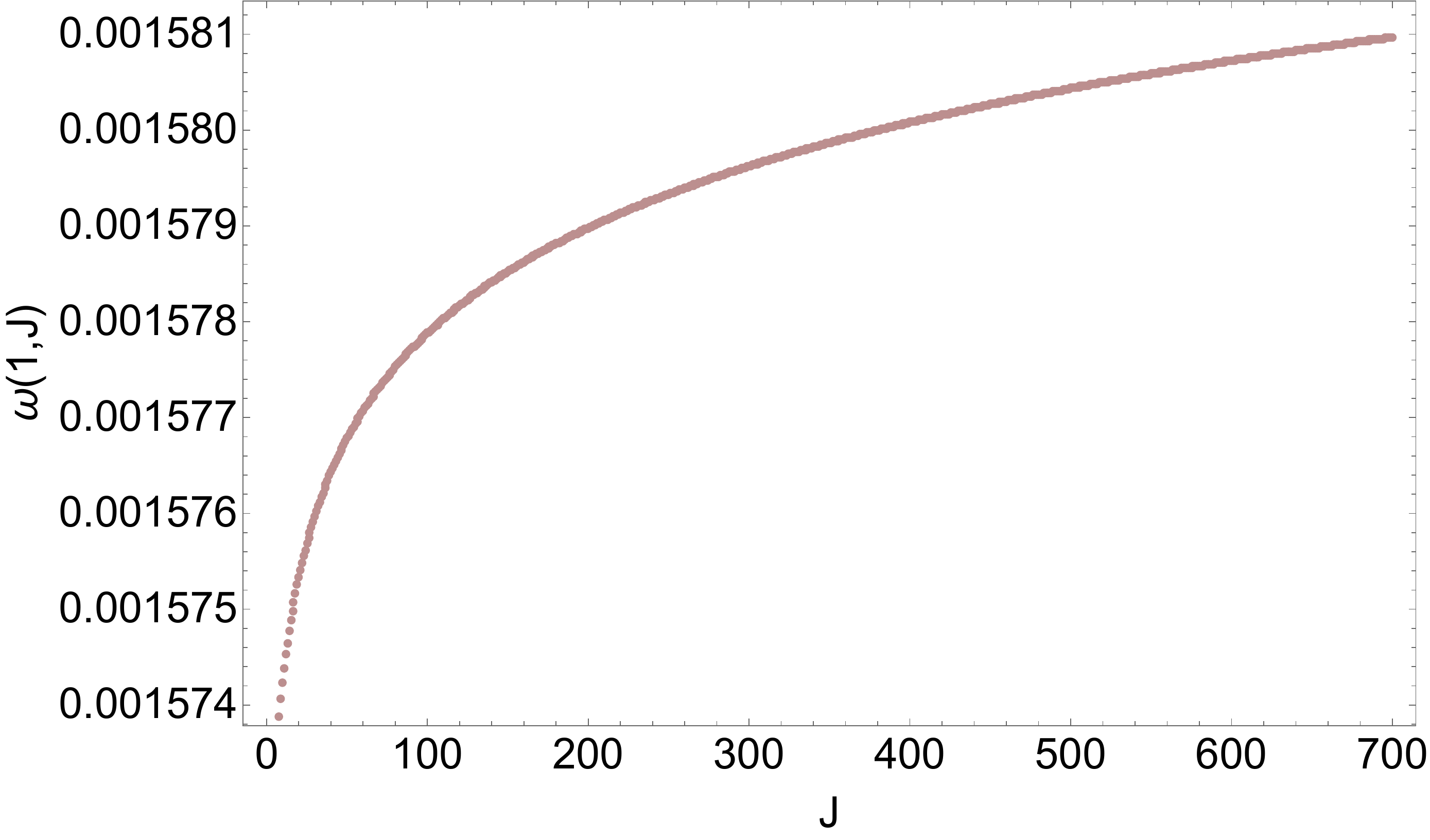}
    \end{subfigure}
    \hfill
    \begin{subfigure}{0.47\textwidth}
    \includegraphics[width=\textwidth]{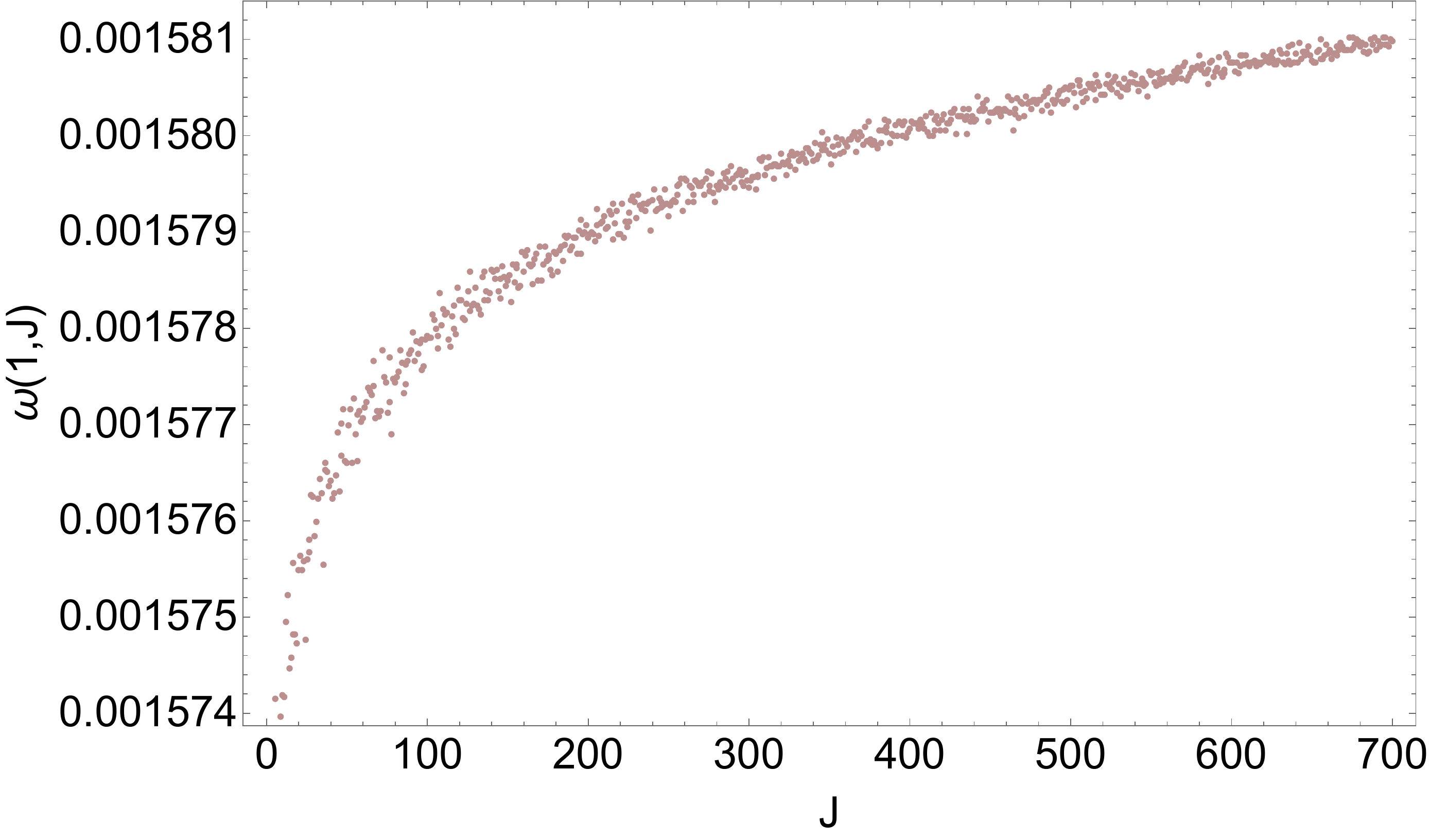}
    \end{subfigure}
    \caption{Spectrum of Rindler with $\sigma_0=0$ (left) vs $\sigma_0=1.0$ (right). We show $\omega(n=1,J)$. } 
    \label{Rin_spectra}
\end{figure}

This therefore again leads to similar structures as in BTZ. We find 
\begin{eqnarray}
 \mu_J = 2\cos(\lambda_J\omega-\theta/2)
\end{eqnarray} 
as well as the quantization condition
\begin{eqnarray}
\cos(\alpha)=\cos(2\lambda_J\omega), \ \ \sin(\alpha)=\sin(2\lambda_J\omega)
\end{eqnarray}

Because the structure is precisely parallel to BTZ, we will not repeat the discussion; it is clear that the normal mode calculation proceeds in an identical manner. The mean value of $\lambda$ can be related to the stretched horizon location. Once we choose $R, \xi_0$ and $a$, the normal modes $\omega(n, J)$ can be numerically solved for as a function of $J$ (and an integer $n$). We present the plots in precise parallel to the BTZ case. The qualitative results are identical, despite the fact that the special functions that showed up in the wave equations here are different. In the plots we present, we have chosen  $a=1, R=2, J_{max}=700, \langle \lambda \rangle = -10^3$ and $\sigma_J=\sigma_o/\sqrt{J}$. The $\sigma_0$ values are quoted in the plots.




\section{The Hairy Harmonic Oscillator and Cut-Off in Empty Space: \\ Level Repulsion without Linear Ramp}

We noted that the  linear ramp in the SFF and repulsion in the LSD can both be seen in the stretched horizon spectrum if the boundary condition is generic. We also pointed out that the level spacing ratio discussed in \cite{Sumilan} is also consistent with RMT expectations. Together, these constitute very strong evidence that fuzzball/stretched horizon spectra have strong connections to random matrices and chaos. 

In this section, we will ask a slightly more resolved question: which of these is a more robust indicator of chaos? Is it the linear ramp or is it level repulsion? Or are both these features always present in systems concomitantly? We will present some hints in this section that the linear ramp may be a more robust diagnostic of strong chaos than nearest-neighbor data. This is not an entirely new suggestion (the length of the ramp is often viewed as an indicator of the ``strength" of chaos), but we will give some examples which we feel are instructive.

\begin{figure}[H]
\begin{subfigure}{0.47\textwidth}
    \centering
    \includegraphics[width=\textwidth]{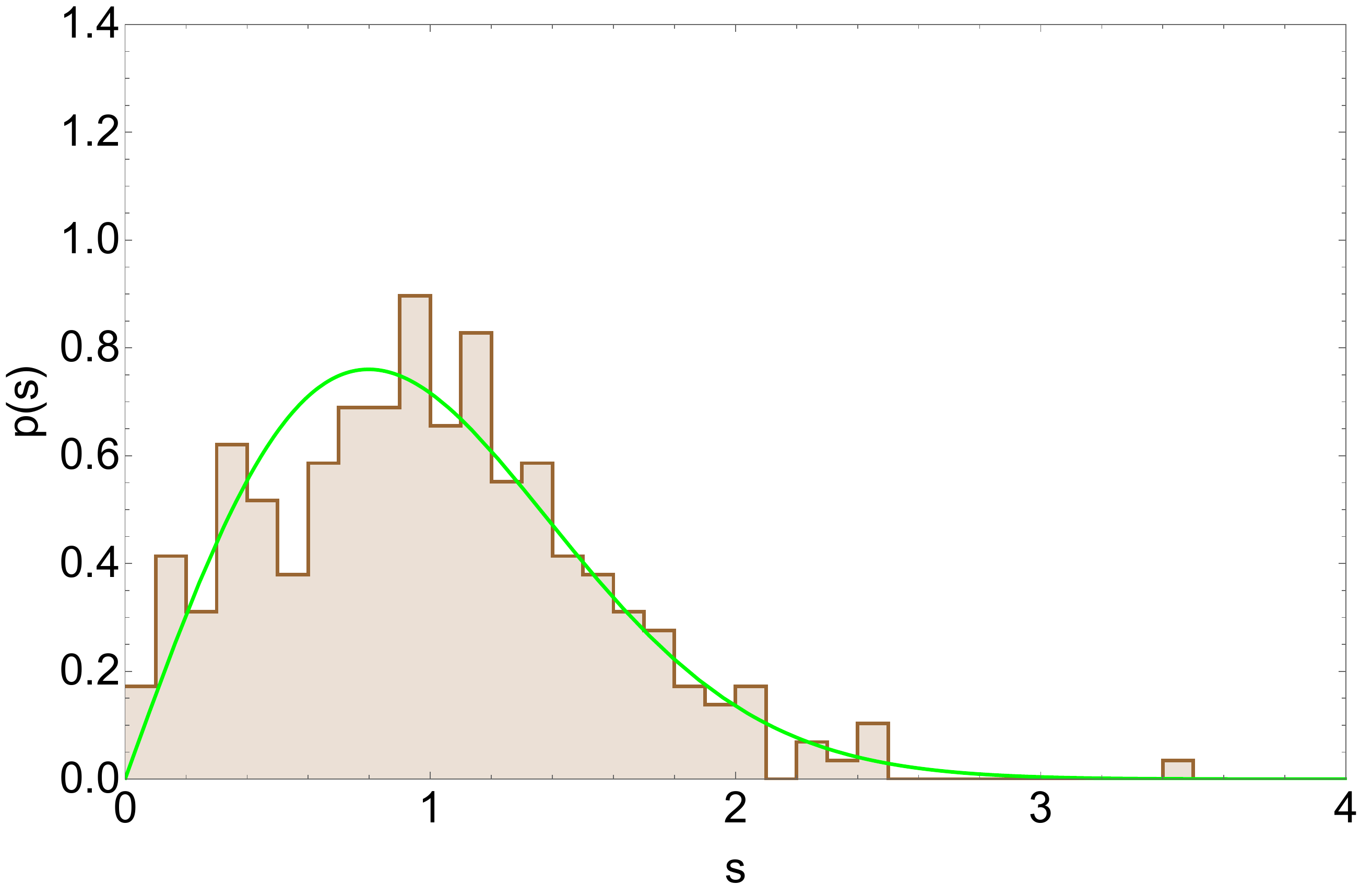}
    \end{subfigure}
    \hfill
    \begin{subfigure}{0.47\textwidth}
    \includegraphics[width=\textwidth]{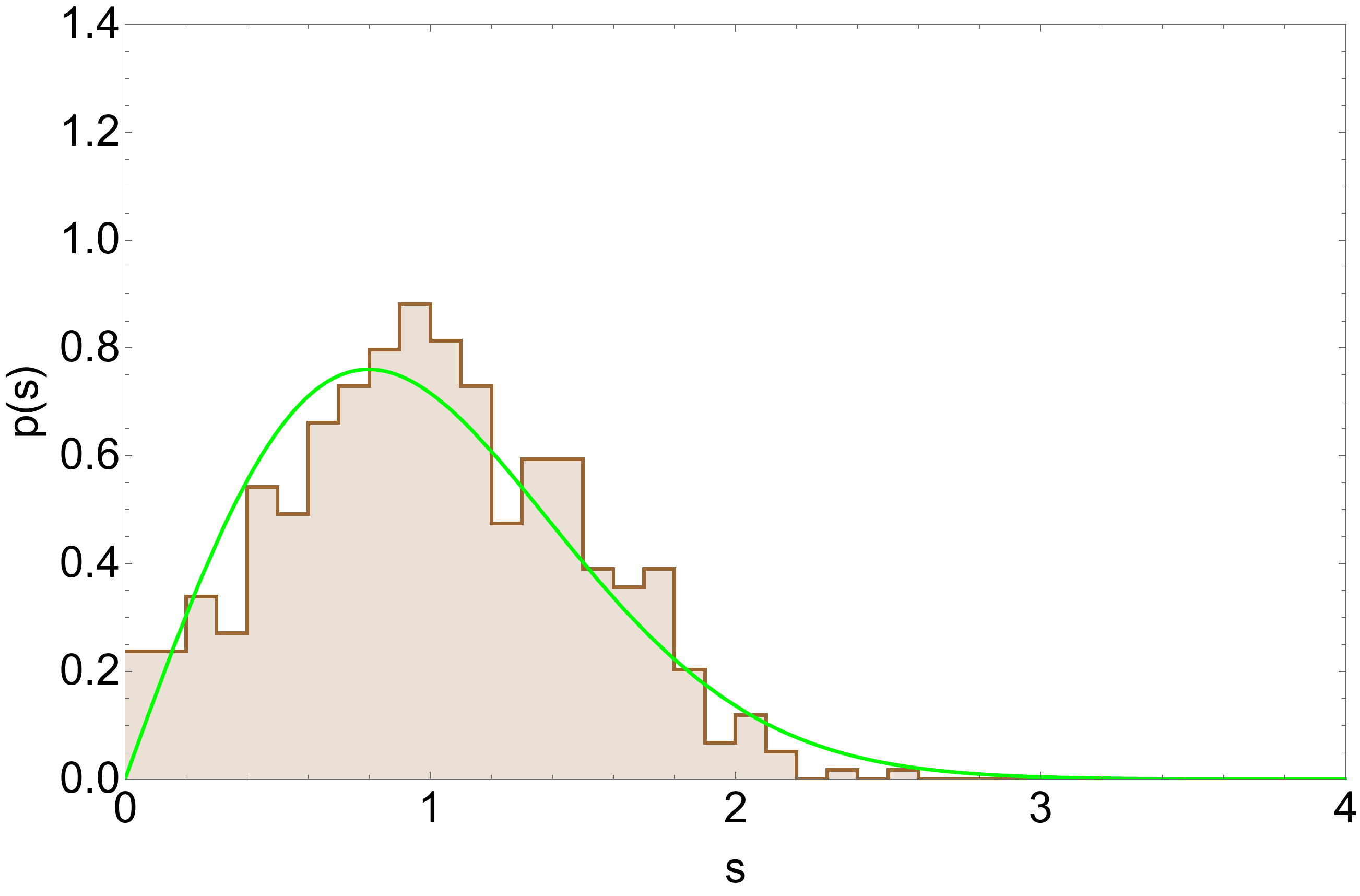}
    \end{subfigure}
    \caption{LSD: Cut-off flat space with fluctuation profile (left) vs hairy SHO (right). Flat space data: $J_{max}=300$, $r_{cut}=1$, $\lambda$-variance $=0.0174$. We are working with $\omega(n=1,J)$. SHO data: $n_{max}=600$, $\omega=1$, spectral noise variance $=0.36$. Both fits are GOE.
} 
    \label{LSD-flat-sho}
\end{figure}

We will start (as often in physics) with the simple harmonic oscillator (SHO). For our purposes, the SHO is interesting because even though it is the farthest thing from a chaotic system, it exhibits a naive (or extreme) version of level repulsion -- the levels are equally spaced, and the LSD is a delta function shifted from the origin. Motivated by the results of this paper, we can ask if there is a natural way to ``perturb" the SHO spectrum so that the level spacing becomes a more conventional Wigner-Dyson-like form. It turns out that a simple way to engineer this exists -- we simply allow a small amount of (Gaussian) noise in the levels of the SHO. We will call this set up a hairy or noisy SHO. See Figure \ref{LSD-flat-sho} right panel, for a typical LSD of an SHO perturbed in this way. We present a GOE fit for concreteness. But again, by adjusting the variance, we can find fits with GSE or GUE. We are not aware of a previous observation of this simple but striking fact in the literature, but it is easy enough to understand -- Random noise in the energy levels directly affects the nearest neighbor data, which explains why the delta function in the LSD gets spread out. (Connecting to our previous results, it also gives a very simple intuition for the fluctuations in the profile functions of fuzzballs -- they are directly responsible for the level repulsion in the microstate spectrum.) 
\begin{figure}[H]
\begin{subfigure}{0.47\textwidth}
    \centering
    \includegraphics[width=\textwidth]{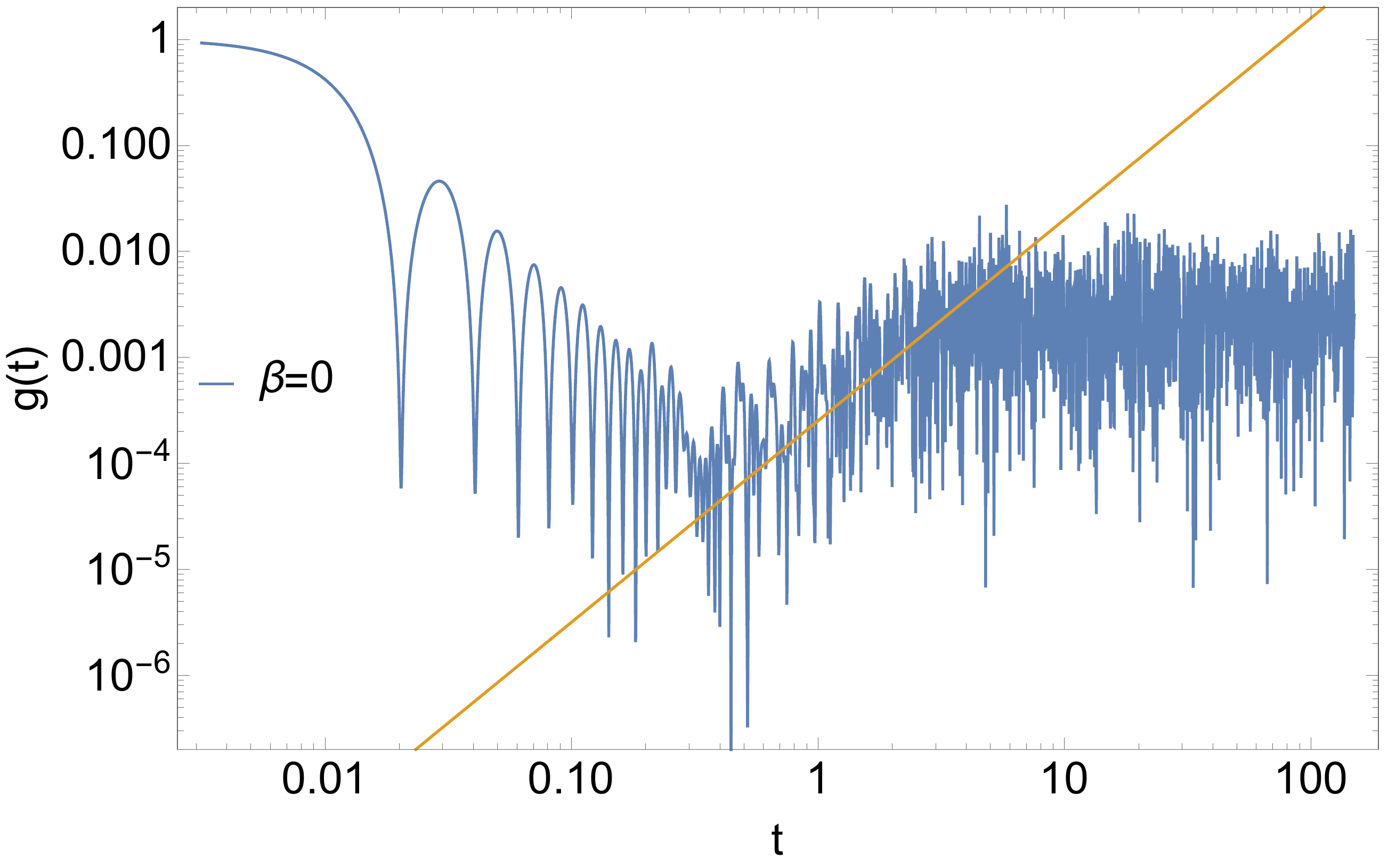}
    \end{subfigure}
    \hfill
    \begin{subfigure}{0.47\textwidth}
    \includegraphics[width=\textwidth]{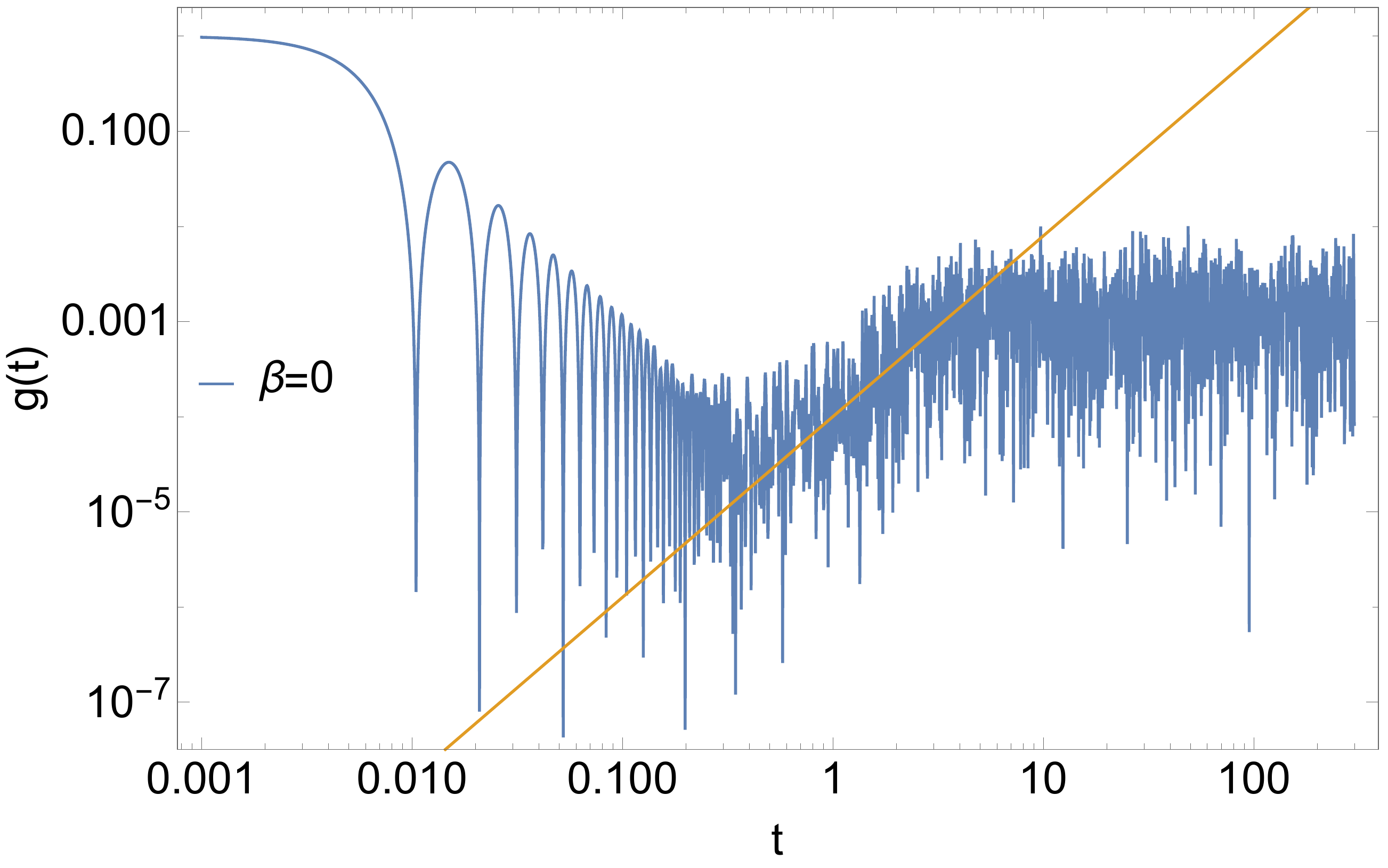}
    \end{subfigure}
    \caption{SFF: Cut-off flat space with fluctuation profile (left) vs hairy SHO (right). 
The data are the same as in the previous plot.     
The yellow line has slope 1.9 (both left and right). In other words, this is a {\em power law} ramp.
} 
    \label{SFF-flat-sho}
\end{figure}
\begin{figure}[H]
\begin{subfigure}{0.47\textwidth}
    \centering
    \includegraphics[width=\textwidth]{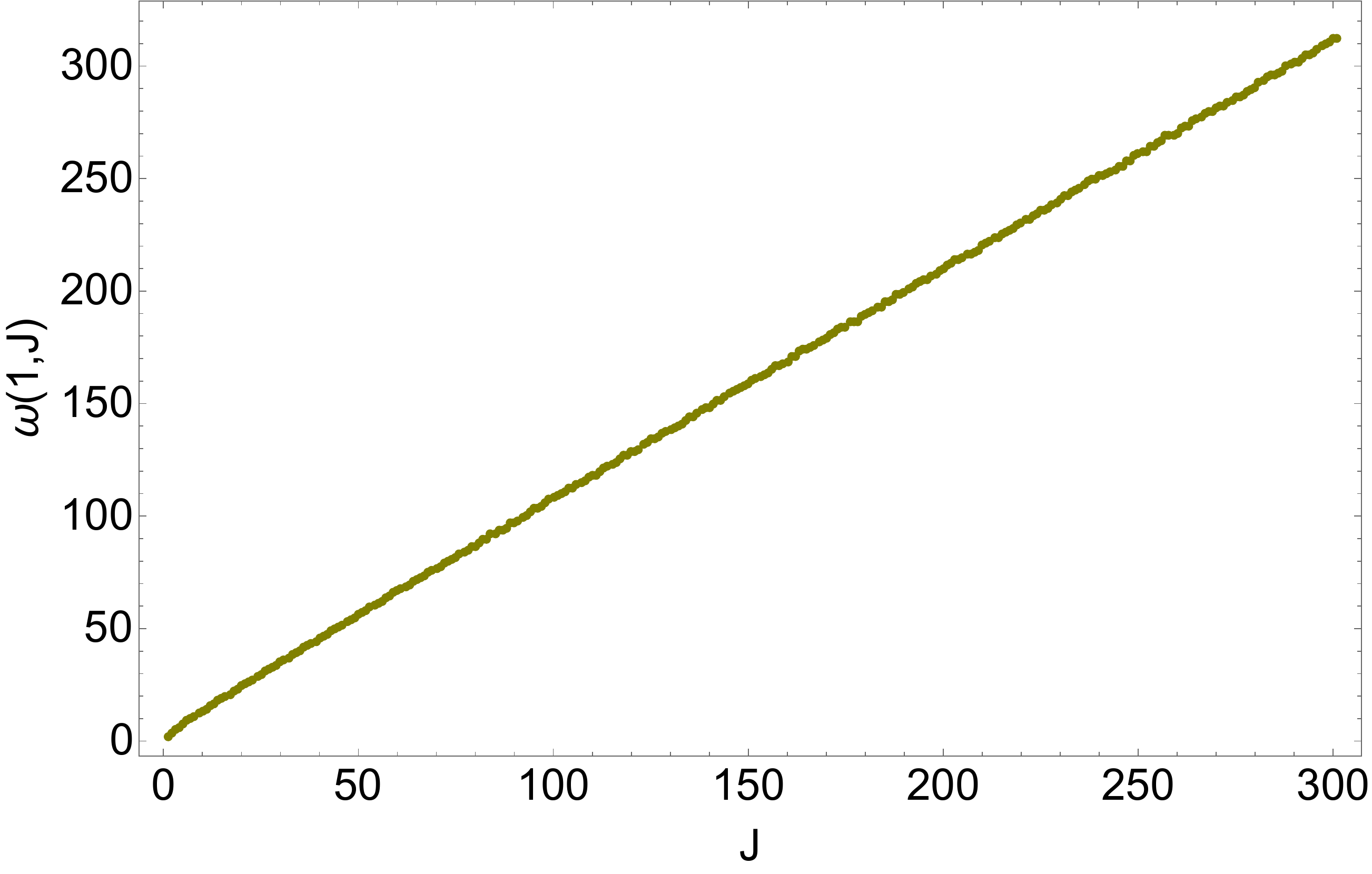}
    \end{subfigure}
    \hfill
    \begin{subfigure}{0.47\textwidth}
    \includegraphics[width=\textwidth]{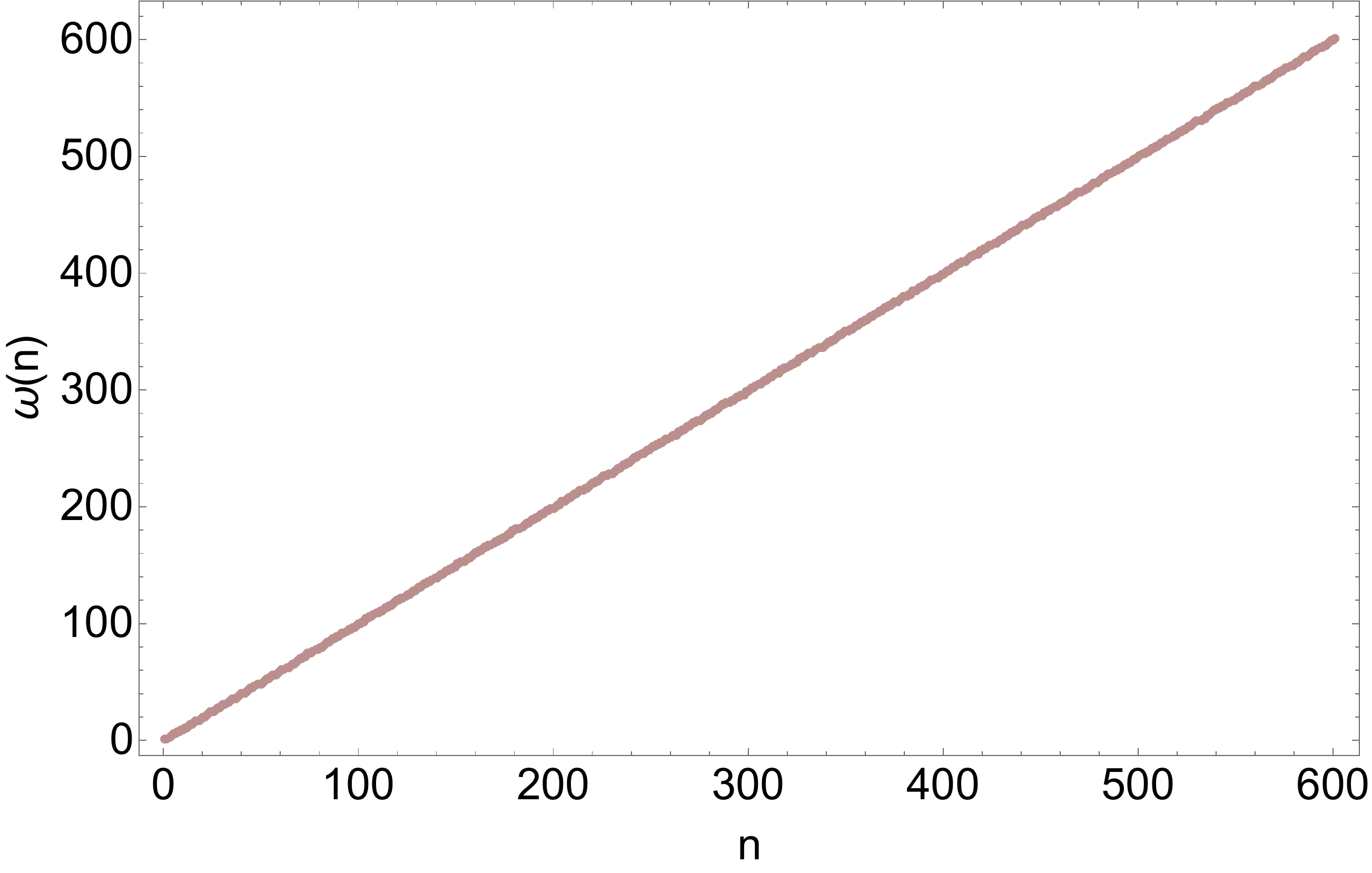}
    \end{subfigure}
    \caption{Shapes of spectra: Cut-off flat space with fluctuation profile (left) vs hairy SHO (right). The data are the same as in the previous two plots. It is clear that both the spectra are approximately evenly spaced. The punchline of the figures in this page is that the spectral features of the two systems have crucial similarities.
} 
    \label{spectrum-flat-sho}
\end{figure}
It is generally believed that strong chaos is characterized by  the linear ramp in the SFF. And indeed, if one computes the SFF of the SHO with noise in the spectrum, one finds that the ramp is in fact non-linear. This is illustrated in Figure \ref{SFF-flat-sho}, right panel. We emphasize that it is remarkable that a well-defined ramp {\em exists}, even though it is not linear. In fact, we find that on a log-log plot, it has a well-defined slope of $\sim 1.9$. In other words, a hairy SHO has a {\em power law} ramp, at least within the context of our numerical results.

\begin{figure}[H]
\begin{subfigure}{0.45\textwidth}
    \centering
    \includegraphics[width=\textwidth]{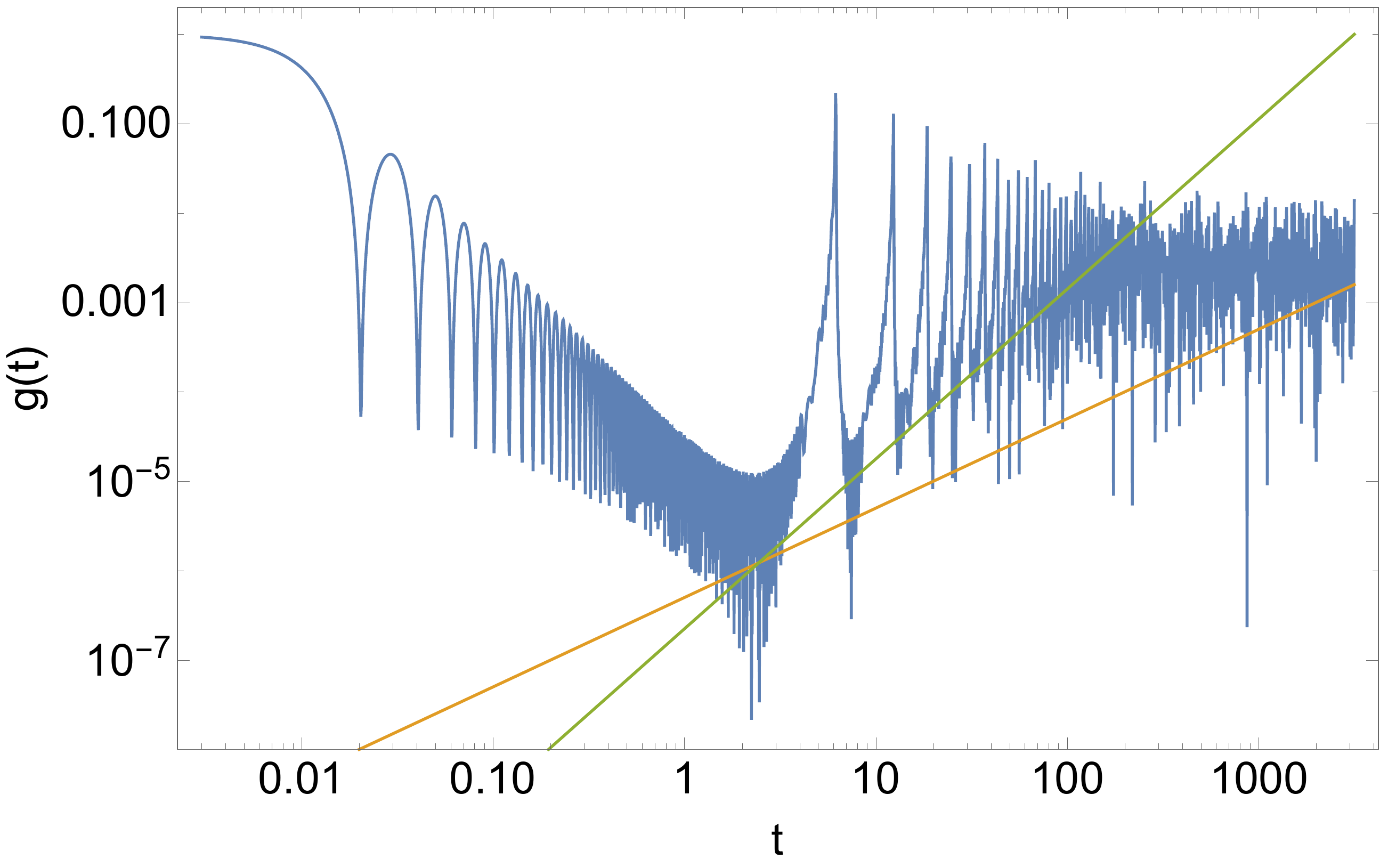}
    
    \end{subfigure}
    \hfill
    \begin{subfigure}{0.45\textwidth}
    \includegraphics[width=\textwidth]{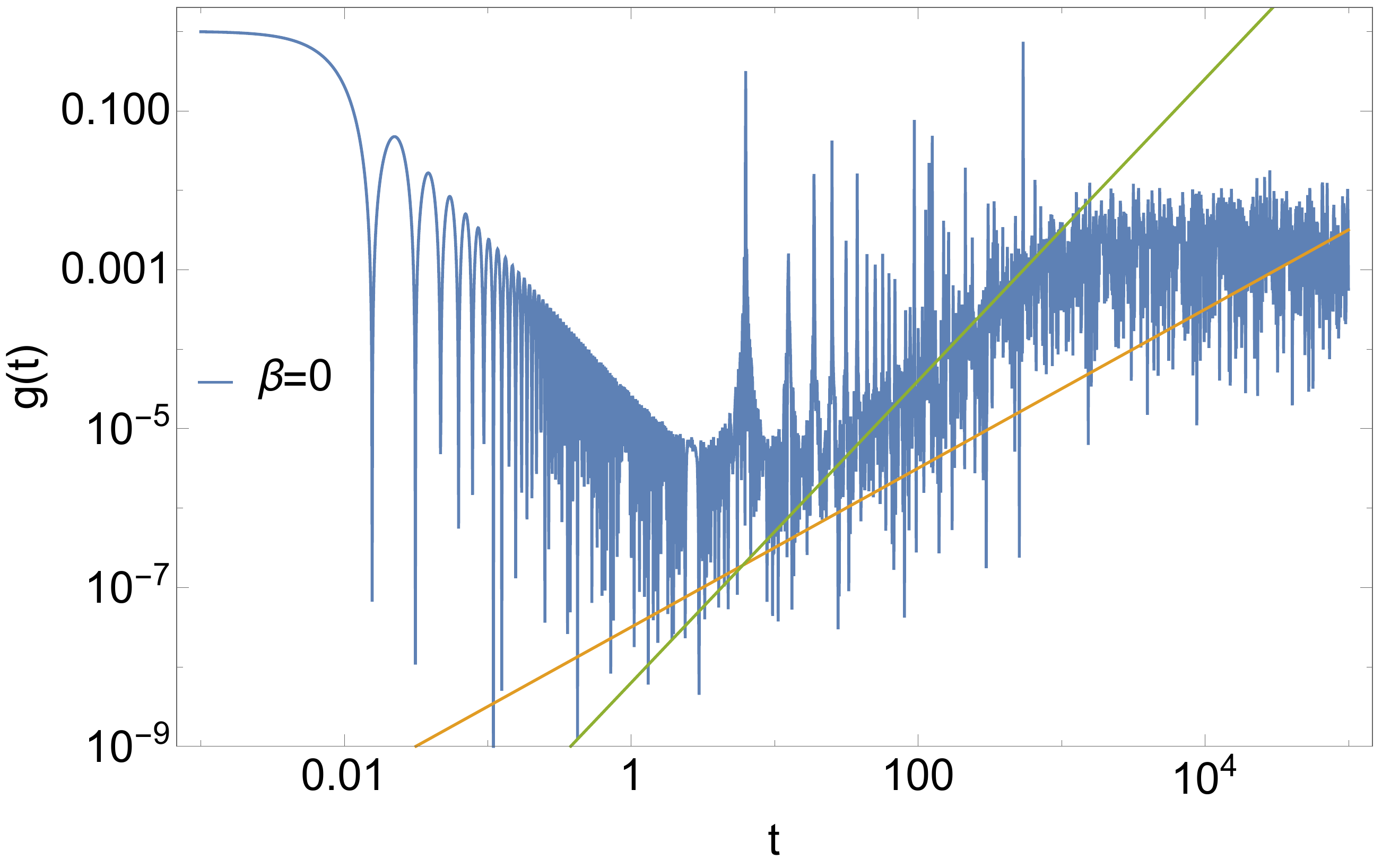}
    \end{subfigure}
    \caption{Cut-off flat space with no variance vs SHO with a tiny amount of noise. The precise values are unimportant. Our goal here is not to make a detailed comparison, but to observe the crude similarity which becomes more striking as we increase the variance/noise. The two lines are of slope $\sim 1.9$ and $\sim 1$.} 
    \label{flat-vs-SHO-extreme}
\end{figure}

These SHO results shed light on the distinctions between  a black hole with a stretched horizon, and a cut-off in empty space. If we impose a simple Dirichlet condition $\phi=0$ at the cut-off, in the former case we find a linear ramp \cite{synthfuzz}, but in empty space there is no clear ramp, certainly nothing of slope $\sim 1$. See Figure \ref{flat-vs-SHO-extreme}. But as we add variance to the profile, we see the emergence of a power law ramp, see Figure \ref{SFF-flat-sho} left panel. The  SHO example above provides us a clear understanding of this. A cut-off in flat space leads to eigenvalues that are connected to the zeros of Bessel functions (as we will see). These are roughly evenly spaced -- so the spectrum looks crudely like that of an SHO.
Relatedly, the level spacing in the $\phi=0$ case is essentially a delta function. But this can be made to look like  a more spread out (WD-like) form by demanding instead that the boundary condition is $\phi \sim \phi_0(\theta)$ where the profile has some variance in its Fourier modes. The noise in the spectrum increases when we do this, and as a result (as pointed out above for the hairy SHO) we find that the LSD takes a more conventional WD form. Of course, when the variance is very large, the spectrum ends up becoming Poisson. Crucially, the slope of the ramp is never $\sim 1$ in these cases. For moderate values of the variance, it is consistent with the $\sim 1.9$ quoted above for the noisy SHO -- see figures. (Note that when the variance is steadily increased, the ramp gets increasingly washed out. So this statement applies only to those values of the variance for which there is a clear ramp.)

The basic message we extract from these calculations is that the spectrum on a cut-off geometry without a horizon is essentially a hairy SHO spectrum. When we have a horizon on the other hand, the spectrum is not that of an SHO in any sense (as we saw in previous sections). Together with the striking linearity of the ramp, we are therefore lead to conclude that the physics in the latter case is {\em not} simply due to nearest-neighbor physics.

We conclude this section by providing some of the details of the flat space calculation. We will work with 2+1 dimensions, the physics we aim for is unaffected by increase in dimensions:
\begin{equation}\label{flat_metric}
    ds^2=-dt^2+dr^2+r^2 d\psi^2
\end{equation}
%
Separating the scalar field as (say) in the BTZ case, we find the radial part
\begin{equation}
   \phi^{''}_{\omega, J}(r)+\frac{1}{r} \phi^{'}_{\omega, J}(r)+   \omega^2  \phi_{\omega, J} (r)
 -V(r)\phi_{\omega, J}(r) =0 \label{radial_eom_flat}
\end{equation}
with \begin{equation}
    V(r)=\frac{1}{r^2}\left( J^2+ m^2 \right).
\end{equation}
We will consider the solution of this equation \eqref{radial_eom_flat} in the massless case, which is given in terms of Bessel functions: 
\begin{equation}\label{sol1-flat}
    \phi(r)= C_1 J_J(\omega r)+ C_2 Y_J(\omega r),
\end{equation}
where, $J_J$ and $Y_J$ are Bessel functions of first and second kind respectively. We suppress the $J$ and $\omega$ (or $n$) subscripts of $C_1$ and $C_2$.

As before, we need one boundary condition to fix a relationship between $C_1$ and $C_2$ and another condition at a cut-off to fix the normal modes. The former role was played by AdS-normalizability in the BTZ case. We could likewise demand a suitably chosen bulk condition here as well that relates $C_1$ and $C_2$. By numerical experimentation  we have found that the qualitative features of the ramp and LSD that we are after, are insensitive to this choice. This is unsurprising because the physics we are interested in, is the result of the quantization condition, and not the relationship between $C_1$ and $C_2$. In the following, we will simply demand that $C_2=0$. Note that this sets the bulk source mode (which is singular at the origin) to zero, while retaining the homogeneous mode. It was noted in \cite{Budhaditya} that the bulk source mode is the analogue in flat space, to the non-normalizable mode in AdS. So this choice is a natural analogue of the normalizability demand in AdS. But we emphasize that large classes of choices are likely to give similar results.

Using this boundary condition, equation \eqref{sol1-flat} becomes
\begin{align}\label{sol3-flat}
   \phi(r)= C_1 J_J(\omega r).\hspace{0.5cm} 
\end{align}
Demanding a profile at the cut-off $r=r_0$ leads to an equations analogous to what we found for BTZ: $\phi(r=r_0)=C_0$. 
\begin{align}
     C_1 J_J(\omega r_0)= C_0 \implies 
        J_J(\omega r_0) = \frac{C_0}{C_1 }\equiv \lambda_J.\label{geneq1-flat}
\end{align}
Note that we could also define the RHS to be $\omega \lambda_J$, which is more analogous to some of our discussions in BTZ and Rindler. But as we mentioned, these choices do not affect the semi-qualitative features we are after, so we will stick with this simple choice here for concreteness.  

We will take $\lambda_J$ to be Gaussian distributed with mean zero, and adjustable variance. The equation is easy to solve numerically, by taking the seed for the root search to be the 1st zero of the $J$-the Bessel function. When the variance is zero, we find an ``extreme'' delta-function like distribution in the LSD. The ramp of the SFF is not particularly well-defined, but we can already see a crude similarity to an SHO with a very small amount of noise -- See Figure \ref{flat-vs-SHO-extreme} below.
When we steadily add variance, we find more conventional level repulsion and the emergence of a robust ramp of slope $\sim 1.9$, which we presented in Figure \ref{SFF-flat-sho} left panel. As noted above, this is precisely what one finds from a noisy SHO as well. Eventually we find a Poisson distributed LSD. The (power law) ramp gets washed out, when the variance becomes very large. These features are identical to what we find in the hairy/noisy SHO case.

To summarize -- flat space with a cut-off is qualitatively identical to hairy SHO. Unlike in the case of the stretched horizon cut-off, the levels are essentially evenly spaced.   
We have done a similar calculation in empty AdS as well, as discussed in the main body of the paper, and the results are again consistent. These results mean that the linear ramp (which is often viewed as an indicator of strong chaos) does {\em not} arise from a cut-off in flat space. But for the same reason that a hairy SHO can mimic the LSD of an RMT (which in itself is a fact {\em not} emphasized previously in the literature, to our knowledge), the spectrum of cut-off flat space can also exhibit level repulsion -- the variance in the boundary condition simply introduces a variance in the nearest neighbor levels. But this is not sufficient to create robust chaos.

A further distinction between empty space with cut-off and the stretched horizon is discussed in the next section.

\section{Planck-scale Hierarchy}

We observed that the fluctuations at the cut-off in empty space translate to fluctuations in the energy levels and therefore lead to level repulsion. In other words, nearest neighbor effects of chaos can be produced simply by having fluctuations at the cut-off.  We also noted however that the linear ramp (which is a deeper signature of chaos) cannot be realized this way, and requires the presence of a horizon. 

In fact there is another interesting distinction between the stretched horizon and a cut-off in empty space. This has to do with the fact that the fluctuations at the cut-off needed in the stretched horizon scenario are hierarchically suppressed, allowing the interpretation that they are Planck-scale. The fluctuations in the empty space cut-off on the other hand are naturally macroscopic. To see this, first note that in \eqref{geneq1-flat}, the first zero of the $J$-th Bessel function is linearly spaced in $J$ with the scale controlled by $r_0$. The natural scale controlling the fluctuations in the RHS is therefore $r_0$ (this dependence is approximately linear if we define the RHS of \eqref{geneq1-flat} to be $\omega \lambda_J$ instead of $\lambda_J$). On the other hand in the  horizon case, the situation is more interesting. To see this in detail, let us work with the concrete case of BTZ, and observe that the conventional tortoise coordinate here is defined via
\begin{eqnarray}
z=\frac{L^2}{2\ r_h} \ln\left(\frac{r+r_h}{r-r_h} \right)
\end{eqnarray}
This means that the usual radial coordinate of the stretched horizon $x \equiv r-r_h$ is approximately 
\begin{eqnarray}
x = 2\ r_h e^{-2r_h z/L^2}, \label{tort2}
\end{eqnarray} 
from which it follows that the fluctuation in the stretched horizon location goes as
\begin{eqnarray}
|\Delta x| \sim 4\ (r_h/L)^2\ e^{-2 r_h z/L^2} |\Delta z| \label{fluctort}
\end{eqnarray}
where we have instated a magnitude sign because $z \rightarrow \infty$ corresponds to the horizon. Now, from \eqref{lambdaeqBTZ} it follows that $e^{2\lambda} = \left(x/r_h\right)$ and therefore 
\begin{eqnarray}
2\ e^{2 \lambda} \Delta \lambda=\frac{\Delta x}{r_h} \implies 2\ x \ \Delta \lambda \sim \Delta x.
\end{eqnarray}
Using  \eqref{tort2} and \eqref{fluctort} in this final relation, we get
\begin{eqnarray}
 \frac{L^2}{r_h} |\Delta \lambda|= |\Delta z|.
\end{eqnarray}
Since the horizon size and AdS length scale are both macroscopic, this means that the fluctuations in $\lambda$ are naturally in tortoise coordinate, implying  via \eqref{fluctort} that the stretched horizon fluctuations are suppressed by a factor of
\begin{eqnarray}
e^{-2r_h z_0/L^2} \label{level-spacing}
\end{eqnarray}
where $z_0$ is the mean stretched horizon in tortoise coordinate. We also see that $L^2/z_0$ is a natural candidate for the Planck length. In units where $L=1$, note that this is a small quantity because $z_0$ is very large when the cut-off is close to the horizon. Of course, since we are working with a fixed background, these are all somewhat heuristic statements.

To summarize: The variance in both cases (with and without horizon) can be used as a heuristic proxy for fluctuations of the cut-off surface. But a key distinction in the stretched horizon is that there, the variance captures the tortoise coordinate and therefore the fluctuations can naturally be viewed as Planck suppressed. 

\section{Open Questions and Future Directions}

In this section, we discuss some questions that are worth understanding better in the wake of our results. Some of these are more conceptual than others.

\begin{itemize}

\item Are there more natural choices for the profile functions? We have considered the most simple-minded notion of a ``generic'' profile -- choose some randomly distributed Fourier coefficients. 
The BPS fuzzball profiles, at least in the 2-charge case  \cite{Rychkov, Avinash} are known to contain enough phase space to reproduce the entropy of the black hole. This suggests that perhaps Haar typicality in some form is a better notion of genericity than our present proposal. It will be interesting to incorporate this in some systematic way. 

\item A remarkable conclusion arising from our calculations is that the fluctuations in the profile functions of fuzzballs are directly repsonsible for level repulsion in the spectrum. In that sense, these profiles geometrize level repulsion.

\item Despite the simplicity of our calculation, we have managed to find a linear ramp with fluctuations and level repulsion in (a heuristic candidate for) a single microstate. The price we have paid is that we have sacrificed a (manifestly) smooth horizon. But the emergence of RMT behavior in our calculation suggests that thermality (and therefore smoothness) may emerge via a suitable ensemble replacement of the microstate. Understanding this operationally is clearly a problem of outstanding interest.

\item In our previous paper \cite{synthfuzz}, the LSD was not one of the conventional RMT distributions, but  there was a clear linear ramp. Our main point in that paper was that this is a generic feature of normal modes at stretched horizons, when the boundary condition $\phi=0$ was imposed. In this paper, we have seen systems which exhibit the opposite behavior -- The ramp is non-linear, but one has level spacing that matches well with conventional Wigner-Dyson-like statistics. In fact, we noticed that the latter can be arranged very simply via an SHO with a noisy spectrum. Together the results of these papers are a very clear demonstration that the folk wisdom that the linear ramp is a smoking gun of conventional Wigner-Dyson classes (or their Altland-Zernbauer generalizations) is {\em not} always true. It will be good to understand the broader setting in which these features arise as special cases. 

\item We did {\em not} have to introduce any form of ensemble average. 
Our profile curve is chosen via a Gaussian distribution in the Fourier coefficients, but it should be emphasized that once the curve is chosen, there is absolutely nothing ``averaged'' about the calculation. The emergence of RMT behavior is entirely deterministic. It has been suggested in \cite{Vyshnav} that semi-classical gravity  should be viewed as a tool for capturing ergodic averaged gravitational dynamics, for evolution that is in bulk local equilibrium. This would give an understanding of the surprising utility of Euclidean gravity in each epoch of Hawking radiation in obtaining the Page curve \cite{Penington}. It will be very interesting to connect these two perspectives. 

\item In \cite{synthfuzz} we had observed that there was a kink-like structure at the top of the ramp. A tangential consequence of the calculations in the present paper is that we have understood that this kink becomes less and less prominent, as we bring the stretched horizon closer and closer to the horizon. This is a strong indication that one of the worries expressed in \cite{synthfuzz} -- that the ramp may be an artefact -- is very unlikely to be true.

\item Inspired by the results of this paper and \cite{synthfuzz}, we have been able to identify a broader class of spectra which lead to interesting ramps and level spacing structures. These results together suggest the notion of a {\em generalized} RMT spectrum, which will be elaborated elsewhere \cite{GRMT}. A key message is that boundary conditions are often a crucial ingredient in quantum chaos. This is true in our black hole problem, but note that the idea is much more general. Eg., the Hamiltonian of the hard sphere gas is simply that of a collection of free particles -- it is the boundary conditions that breathe life (and chaos) into the system. 

\item One of the technical features underlying the results of this paper and \cite{synthfuzz} is the observation that the dependence of the spectrum on the angular quantum numbers is not linear. Instead it gets pulled logarithmically along $J$. The resulting quasi-degeneracy was essential for our results. It will be good to get a more mechanical/conceptual understanding of this observation as well as to explore its consequences more broadly. 

\item We found a clear ramp with slope $\sim 1.9$ in our SFF plots for hairy SHO and cut-off flat space. This is an extremely simple example of a non-linear ramp, whose slope is a constant ($\neq 1$) in a log-log plot. It seems surprising and interesting that it is closely related to the SHO. Can this shed light on the fact that despite being the ``ultimate'' integrable system, the SHO exhibits an extreme version of level repulsion (ie., its LSD has no support at the origin, and has a delta function form)?

\item Relatedly, and more speculatively -- does the fact that extreme WD spectra arise from Dirichlet boundary conditions at stretched horizons indicate that black holes are the ``ultimate'' RMT systems? If this is true, black holes can be viewed as the natural counterpoint to SHOs from our previous item. Note that the suggestion we are making here is distinct from the chaos bound of \cite{MSS}, which is about early time chaos and OTOCs. The observation about LSDs that we are making here is related to late time chaos. Black holes may not just be fast scramblers \cite{Sekino}, they may also be the most {\em robust} scramblers. Clearly, more work remains to be done.


\end{itemize}

\begin{thebibliography}{99}
\bibitem{linear} In this paper, by linear ramp we will mean a ramp of slope $\sim 1$ on the log-log plot. A constant slope ramp on the log-log plot, but with a slope different from unity, is still non-linear.
\bibitem{SFF} See \cite{Cotler} for the general definition of SFF and \cite{synthfuzz} for discussions on it in our context. We always work at infinite temperature, $\beta=0$, in this paper. LSD is defined and discussed in \cite{Haake}.
\bibitem{rampy-spectra} We have since been able to construct many examples of this type, this will be presented elsewhere \cite{GRMT}.
\bibitem{LSR0} Even though  we do no report the details here, we have also studied the level-spacing ratios $\gamma$ \cite{Sumilan} of these spectra. This is another diagnostic of RMT behavior along with SFF and LSD. For small/zero variance, we find $\gamma$ values that are consistent with RMT spectra. But it steadily increases with the variance and becomes (very) large, matching the expectation that $\gamma=\infty$ for Poisson systems \cite{Sumilan}.  $\gamma$ is a diagnostic defined via nearest neighbor data and is therefore somewhat redundant with the LSD. This is one reason why we do not consider this as truly distinct diagnostic, and do not emphasize it in this paper. In all the examples we consider, the behavior of LSD and LSR are mutually consistent. The LSD and the (linear ramp of the) SFF on the other hand, do genuinely capture somewhat distinct aspects of random matrix behavior as we will elaborate.
\bibitem{uniform} The precise distribution does not seem too important for our results. This is natural because (as noted in our motivations), we are looking for results like linear ramp and level repulsion, which are semi-qualitative and robust. We have checked that similar statements hold also for uniformly distributed Fourier modes, but we will not elaborate on it here.
\bibitem{WhichWD} By choosing the variance suitably, we can get good fits with GSE, GUE or GOE. We will mostly present GUE fits in this paper. A very interesting feature of these results is that since they arise by tuning certain continuous boundary conditions and not the (discrete choice of) ensemble from which the Hamiltonian matrix is chosen, they seem to allow a continuum  of LSDs that naturally generalize WD.
\bibitem{fluc} 
Let us also re-iterate that the fluctuations should naturally be viewed as macroscopic (and not Planck suppressed) if they are to give rise to level repulsion in a cut-off geometry without a horizon.
\end{thebibliography}
\end{document}